\documentclass[12pt,letterpaper]{article}
\usepackage{epsfig,setspace,amsmath,epsf,amssymb,bm,theorem,cite,graphicx, epstopdf,algorithm,algpseudocode,float,color,mathtools,authblk,tikz,physics}
\usetikzlibrary{positioning}
\usetikzlibrary{decorations.text}
\usetikzlibrary{decorations.pathmorphing}
\usepackage{booktabs}
\usepackage{arydshln}
\usepackage[short,c2]{optidef}
\usepackage{enumitem}
\usepackage{subfigure}
\usepackage{xcolor}

\newtheorem{theorem}{Theorem}

\newtheorem{definition}{Definition}
\newtheorem{remark}{Remark}

\newenvironment{Proof}[1]{\medskip\par\noindent{\bf Proof:\,}\,#1}{{\mbox{\,$\blacksquare$}\par}}

\allowdisplaybreaks

\title{Multi-Threshold AoII-Optimum Sampling Policies for CTMC Information Sources\thanks{This work is presented in parts at IEEE Infocom 2024 and IEEE ISIT 2024. This work is done when N.~Akar is on sabbatical leave as a visiting professor at University of Maryland, MD, USA, which is supported in part by the Scientific and Technological Research Council of T\"{u}rkiye  (T\"{u}bitak) 2219-International Postdoctoral Research Fellowship Program.}}

\author[1]{Ismail Cosandal}
\author[2]{Nail Akar}
\author[1]{Sennur Ulukus}

\affil[1]{\normalsize University of Maryland, College Park, MD, USA}
\affil[2]{\normalsize Bilkent University, Ankara, T\"{u}rkiye}

\begin{document}
\date{}
\maketitle

\vspace*{-1.0cm}

\begin{abstract}
We study push-based sampling and transmission policies for a status update system consisting of a general finite-state continuous-time Markov chain (CTMC) information source with known dynamics, with the goal of minimizing the average age of incorrect information (AoII). The problem setting we investigate involves an exponentially distributed delay channel for transmissions and a constraint on the average sampling rate. We first show that the optimum sampling and transmission policy is a \emph{multi-threshold policy,} where the thresholds depend on both the estimation value and the state of the original process, and sampling and transmission need to be initiated when the instantaneous AoII exceeds the corresponding threshold, called the estimation- and state-aware transmission (ESAT) policy. Subsequently, we formulate the problem of finding the thresholds as a constrained semi-Markov decision process (CSMDP) and the Lagrangian approach. Additionally, we propose two lower complexity sub-optimum policies, namely the estimation-aware transmission (EAT) policy, and the single-threshold (ST) policy, for which it is possible to obtain these thresholds for CTMCs with relatively larger number of states. The underlying CSMDP formulation relies on the \emph{multi-regime phase-type} (MRPH) distribution which is a generalization of the well-known phase-type distribution, which allows us to obtain the distribution of time until absorption in a CTMC whose transition rates change with respect to time in a piece-wise manner. The effectiveness of the proposed ESAT, EAT and ST sampling and transmission policies are shown through numerical examples, along with comparisons with a baseline scheme that transmits packets according to a Poisson process in out-of-sync periods.  
\end{abstract}

\section{Introduction} \label{sec:int}
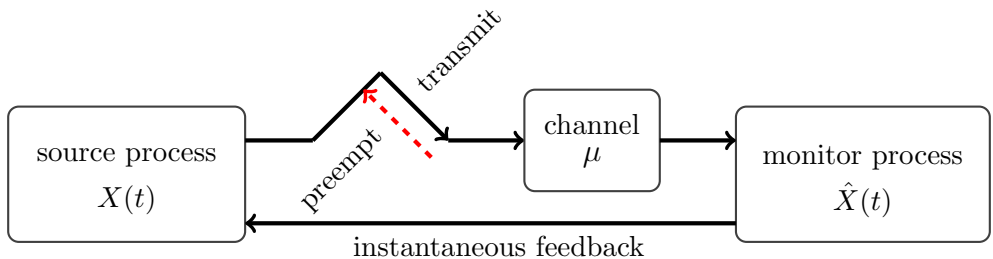
\begin{figure}[bht]
    \centering
    \begin{tikzpicture}[scale=0.45]
    \draw[rounded corners,thick,darkgray] (0,-1) rectangle (7,3) {};
    \filldraw (3.5,1.5) node[anchor=center] {\small{source process}};
    \filldraw (3.5,0.25) node[anchor=center] {\small{$X(t)$}};
    \draw[rounded corners,thick,darkgray] (21.5,-1) rectangle (29,3) {};
    \filldraw (25.25,1.5) node[anchor=center] {\small{monitor process}};
    \filldraw (25.25,0.25) node[anchor=center] {\small{$\hat{X}(t)$}};
    \draw[ultra thick,->,black] (21.5,-0.45) -- (7,-0.45);
    \filldraw (14.5,-0.50) node[anchor=north] {\small{instantaneous feedback}};
    \draw[ultra thick] (7,2) -- (9,2);
    \draw[ultra thick] (9,2) -- (11,4);
    \draw[->,black,ultra thick] (11,4) -- (13,2);
    \node at (11,2) (nodeA) {};
    \node at (16,7) (nodeB) {};
    \draw[ultra thick,->] (13,2) -- (15.25,2);
    \draw [decoration={text along path,
    text={|\small|transmit},text align={center}},decorate]  (nodeA) -- (nodeB);  
    \draw[->,red,ultra thick,dashed] (12.5,1.5) -- (10.5,3.5);
    \node at (8,-1) (nodeA) {};
    \node at (12,3) (nodeB) {};
    \draw [decoration={text along path,
    text={|\small|preempt  },text align={center}},decorate]  (nodeA) -- (nodeB);
    \draw[rounded corners,thick,darkgray] (15.25,0.5) rectangle (19.25,3.5) {};
    \filldraw (17.25,2.5) node[anchor=center] {\small{channel}};
    \filldraw (17.25,1.5) node[anchor=center] {$\mu$};
    \draw[ultra thick,->] (19.25,2) -- (21.5,2);
    \end{tikzpicture}
    \caption{A remote estimation system with the source process $X(t)$ and the monitor process $\hat{X}(t)$ for which the source employs a certain transmission policy to transmit the status update packets via a delay channel, and also a preemption policy to preempt ongoing transmissions when the observed information becomes obsolete.}
    \label{fig:SystemModel}
\end{figure}

We consider the remote estimation of a finite-state irreducible 
continuous-time Markov chain (CTMC) information source process $X(t) \in \mathcal{N} = \{1,2,\ldots,N\}$, and a remote monitor process $\hat{X}(t)$ with the same state-space. The remote monitor and the source are said to be in-sync at time $t$ when $\hat{X}(t)=X(t)$, and out-of-sync at time $t$ when $\hat{X}(t) \neq X(t)$. In order to provide synchronization, the source's sensor has the capability to initiate a sampling operation at any time with the sampled values immediately sent by the source's transmitter to the remote monitor encapsulated in information packets, over a delay channel. For the sake of brevity, the joint sampling and transmission operation is referred to as \emph{transmission} throughout the paper. The remote monitor estimates the state of the source process with the value of the sample encapsulated in the latest received information packet. We assume perfect and instantaneous feedback and the source is therefore always aware of transmission completion instances. Preemption of a packet in transmission is enforced when the state of the source changes during an ongoing transmission. The system is illustrated in Fig.~\ref{fig:SystemModel}.

In this paper, we employ the \emph{age of incorrect information} (AoII) metric proposed in \cite{maatouk2020} which penalizes the mismatch between the source and the monitor with the time elapsed since they have become out-of-sync, but the proposed framework is suitable for more general AoII penalty functions as well. 

Imposing a constraint on the average sampling rate which bounds the frequency of packet transmissions, we study three transmission policies with the goal of minimizing the average AoII of the status update system. In the \emph{estimation- and state-aware transmission} (ESAT) policy, the source initiates a transmission once AoII exceeds a threshold depending on the instantaneous values of both the source and the monitor processes. Note that AoII is zero throughout an in-sync period during which there is no need for a transmission. The ESAT policy is shown in this paper to be optimum with the proper choices of $N(N-1)$ thresholds. However, obtaining these thresholds may not be computationally feasible for large values of $N$. By means of relaxation of the ESAT problem, we propose a sub-optimal \emph{estimation-aware transmission} (EAT) policy for which the threshold values are allowed to depend on the estimation values only. EAT requires the calculation of $N$ thresholds for which we propose a computationally efficient algorithm with lower complexity that employs one-dimensional search, i.e., line search, at a given iteration of the algorithm. The sub-optimal EAT policy gives rise to the optimum transmission policy for binary (two-state) CTMC sources. The final transmission policy studied in this paper is the \emph{single-threshold} (ST) policy for which transmission is initiated once the single system-wide threshold is exceeded, irrespective of the instantaneous values of the source and the monitor processes. We show that the optimum threshold value for this policy can be obtained easily by line search. We also note that ST provides the optimum policy for symmetric CTMCs. 

When any one of the three studied policies is employed along with given thresholds, we model the underlying status update system with an embedded discrete-time Markov chain (DTMC) of size $N$ that makes state transitions at synchronization points at which the monitor just synchronizes with the information source. Synchronization can stem from either with a completed transmission, or from a certain change of the source state. For a given threshold policy, and for each  state of the embedded DTMC, we find  (i) state transition probabilities, (ii) average total accumulated AoII (area under the AoII curve), and (iii) average number of transmissions, when this particular state of the embedded DTMC is visited, using the theory of \emph{phase-type} (PH-type) distributions which stand for the distribution of time until absorption in a CTMC with an absorbing state \cite{neuts81, latouche1999introduction}. In this paper, we develop a generalization of PH-type distributions in two directions: (i) we introduce multiple absorbing states, (ii) we allow the generator of the CTMC to change in a piece-wise constant manner as a function of elapsed time, governed by the multiple thresholds of the underlying transmission policy.  Inheriting the terminology from \cite{kankaya_akar_sm08}, we call the proposed distribution \emph{multi-regime phase-type} (MRPH) by defining a number of finite intervals (or regimes) in time during each of which the generator of the underlying CTMC is intact, but it may change when we transition from one regime to another. We derive the absorption probabilities, and also the first two moments of the time until absorption in this paper for MRPH distributions, which enables us to exactly analyze the DTMC between two synchronization points. 

For the ST policy, we find the optimum threshold using the analytical method described above, in conjunction with line search. For the ESAT and EAT policies involving multiple thresholds, we model the problem as an infinite-horizon constrained semi-Markov decision process (CSMDP) since the \emph{sojourn times} between two successive synchronization points are generally distributed. 
Moreover, the state-space of the proposed CSMDP consists of $N$ states only, corresponding to the synchronization values at synchronization instances. An apparent advantage of the proposed CSMDP is that the thresholds constitute the continuous action space, and the AoII values are not a part of the state-space description of the CSMDP, avoiding a need for neither truncation nor discretization of the state-space. Subsequently, we convert the CSMDP to a standard-form for a semi-Markov decision process (SMDP) without constraints, by employing the Lagrangian method adapted from \cite{white1993mdp, ibe2013markov, hu2007markov}, and then obtain optimum policies using the policy improvement algorithm, given for example in \cite{makowski1986}.

Throughout the paper, certain minimization problems are reduced to line search where the optimum value is obtained by minimizing a function of a single parameter. We particularly use \emph{bisection search} and \emph{gradient descent} algorithms to find the value of the minimizing parameter. By bisection search \cite{AppliedNumericalAnalysis}, we refer to an interval search algorithm in which the search interval is halved at each iteration by evaluating the constraint function at the midpoint of the interval. This method helps us to find the boundary of the constraint set if it is a convex set. The gradient descent algorithm, on the other hand, refers to searching the minimum point in the opposite direction of the derivative of the function when it is available, and it is known that a quasi-convex problem can be solved with gradient descent \cite{boyd2004convex}. We refer the reader to \cite{AppliedNumericalAnalysis}, \cite{floudas2008encyclopedia} and \cite{lu2022gradient} for other alternatives for line search. 

The contributions of this paper can be summarized as follows:
\begin{itemize}
    \item While the average AoII metric is generally investigated in discrete-time settings and for special Markov chains only, to the best of our knowledge, ours is the first paper (along with our previous works in \cite{cosandal2024modelingC, cosandal2024aoiiC}) to consider finite-state irreducible Markov chains in continuous-time in this generality from the AoII perspective.
    \item We propose the generalized version of PH-type distribution, namely MRPH distribution, which allows us to find the distribution of time before absorption in a CTMC when the underlying generator of the CTMC changes in time in a piece-wise constant manner. This new tool is promising in helping solve similar multi-threshold problems in the literature. 
    \item Sub-optimal single-threshold policies are widely used in the AoII literature with optimality proven only for special cases. In our work, we prove that the optimum transmission policy for a general CTMC source is a multi-threshold policy, called ESAT, for which the threshold values depend on both the current estimation and the current state of the source. We propose an $N$-state CSMDP formulation with continuous action space for ESAT making use of the MRPH distribution we develop for the first time in this paper, while employing multi-dimensional search over $N-1$ thresholds in the policy improvement step. We reduce the constrained problem to the unconstrained problem by using the Lagrangian approach, and obtain the optimum Lagrangian coefficient with the bisection method.
    \item We additionally propose two lower complexity policies namely EAT and ST, which are shown to be optimum for special CTMCs while we provide closed-form expressions for the corresponding thresholds in these special cases.
    \item We obtain the threshold for the ST policy for the general CTMC case using an algorithm with computational complexity $O(N^4)$ involving an analytical method accompanied by line search. Separately, we obtain the thresholds for the EAT policy using the CSMDP formulation, and the gradient descent algorithm in the policy improvement stage, with a computational complexity of $O(N^4)$ per policy improvement. Hence, the optimum EAT and ST policies can be obtained numerically for relatively large state-space CTMCs owing to the exclusion of instantaneous AoII in the state-space description of the underlying CSMDP.    
\end{itemize}

The remainder of the paper is organized as follows. In Section~\ref{sec:rw}, related works from the literature are presented. In Section~\ref{sec:PD}, PH-type and multi-regime PH-type distributions are described. The system model is presented in Section~\ref{sec:sys}. In Section~\ref{sec:esat}, the optimization problem is formulated, and ESAT is introduced, whereas Section~\ref{sec:relaxed} is dedicated to the analysis and optimization of EAT and ST policies. Numerical results are presented in Section~\ref{sec:nr}, and conclusions are given in Section~\ref{sec:conc}.  

\section{Related Work} \label{sec:rw}
The freshness of the received information plays a crucial role in remote estimation problems. To quantify information freshness, AoI has been proposed in \cite{Yates__HowOftenShouldone}, which is an indicator of how long ago the latest arrived information in the system has been generated. Following this work, several extensions of this metric have been proposed which include but are not limited to version age of information \cite{bastopcu2020should}, value of information \cite{ayan2019value}, age of synchronization (AoS) \cite{tang_etal_twc20}, and age of incorrect information (AoII) \cite{maatouk2020}. Among these freshness metrics, AoII is of particular interest since it considers the dynamics of the source process, and thus it is considered as a semantic metric \cite{lu2022semantics, maatouk2022age}. It is proposed in \cite{maatouk2020} as a way of quantifying the age of mismatch between the source and the remote monitor, or the \emph{staleness} of the information. The main motivation behind this metric is that even though the latest received packet may have been generated a long time ago, it is possible that the source may not have changed since then, and therefore, the packet can still be fresh. Similarly, a recently received packet may contain stale information, if the source has already changed its state after the packet was generated. Another interesting feature of AoII in contrast to AoI is that the monitor is not required to get a new sample to reset the age since the disagreement between the source and the monitor may come to an end with a transition of the source to the estimated value at the monitor. This metric also differs from conventional mismatch metrics such as the mean square error (MSE), binary freshness metric \cite{bastopcu2021, cosandal2023timely}, or Hamming distortion measure \cite{imer2005optimal}, because the penalty increases as the mismatch remains.

There are several freshness metrics proposed in the literature that are inspired by AoII. Age penalty is defined as the time after the latest state transition in \cite{champati2022detecting}. Different from AoII, the age penalty is not reset when the source process returns to the estimated value. In \cite{joshi2021minimization}, age of incorrect estimation (AoIE) is proposed for an auto-regressive Markov process where the estimate is accepted as the correct estimate while the estimation error is below a certain value, and the value of the AoIE increases with the cumulative sum of the estimation error while the estimation is not correct. In \cite{salimnejad2024version}, the age of incorrect version (AoIV) is proposed that keeps track of the number of state changes at the source while the source and the monitor are out-of-sync.

The AoII metric is mostly investigated in a discrete-time setting. In \cite{maatouk2020, kam2020age, maatouk2022age, chen2022preempting, chen2021minimizing,delfani2024optimizing}, push-based system models with DTMC information source processes are studied. In these works, optimum transmission policies are obtained so as to minimize the average AoII with and without sampling rate constraints, in the form of a threshold policy, and the optimization problem is modeled as a Markov decision process (MDP). However, only special DTMCs are studied in these works. In particular, references \cite{maatouk2020} and \cite{kam2020age} study binary sources, whereas \cite{maatouk2022age,chen2022preempting,delfani2024optimizing} study DTMCs with a general number of states, but they are assumed to be symmetric. In \cite{delfani2024optimizing}, an MDP problem with two constraints is converted to an unconstrained MDP with a token-based approach for both AoI and AoII minimization problems. Reference \cite{chen2021minimizing} studies a source process modeled as a birth-death Markov process. For DTMC sources, there may not always be a single optimum threshold, but it may be a mixture of two thresholds \cite{bountrogiannis2023age, maatouk2022semantics}. Different from the aforementioned works, in \cite{kriouile2022pull}, a pull-based system model with a symmetric DTMC source is studied. Since the scheduler does not know whether its information is correct or not in this system model, the problem is modeled as a partially observable MDP (POMDP). Even though the authors of \cite{kriouile2022pull} show that the problem itself is a restless multi-armed bandit problem (RMAB), they propose a low-complexity solution using Whittle's index and the Lagrangian relaxation approach. The authors of \cite{saha2022relationship}, on the other hand, consider a piece-wise linear source model which takes  continuous values, and they investigate the relationship between AoII and MSE.

Threshold policies also play a crucial role in the AoI literature.  In \cite{arafa2019age}, an energy harvesting sensor with a finite-sized battery is considered, and the AoI-optimum energy-efficient transmission policy is investigated. In this work, it is shown that the optimum sampling strategy is a \emph{threshold} policy for which the sensor transmits only when the AoI exceeds a threshold, and the value of the threshold depends on the energy value in the battery. Even more interestingly, it is shown that in the presence of channel delay, transmitting after waiting is better than zero-delay policies even without considering sampling costs \cite{kam2020age, li2020waiting, bedewy2019age}. In \cite{kam2020age, li2020waiting}, the optimum AoI-minimizing policy is shown to be a threshold policy. In \cite{bedewy2019age}, a single source with multiple servers is considered and the optimum number of responses is obtained before updating the status. Similarly, in \cite{atabay2020improving}, a multiple access scenario is studied, and the authors proposed a threshold policy, namely lazy ALOHA. This study also does not consider an energy or sampling constraint, and they show that AoI performance of lazy ALOHA is better than {slotted ALOHA}.

In our previous works \cite{cosandal2024modelingC, cosandal2024aoiiC}, we investigate the AoII performance of a CTMC source under sampling constraints. In \cite{cosandal2024modelingC}, we separately consider push-based, and pull-based system models. In the pull-based model, the monitor sends pull requests based on the estimation value of the monitor, and the sources start transmission after receiving these requests. On the other hand, in the push-based model, the source starts the transmission after a random threshold value is exceeded.  In \cite{cosandal2024aoiiC}, only the push-based model is investigated but in more detail, and we provide a CSMDP formulation of the EAT policy. In the current paper, we extend the work in \cite{cosandal2024aoiiC} to cover also the ESAT policy and we show its optimality.

There are also other recent works that consider CTMCs in the remote estimation literature. In \cite{inoue2019aoi}, the probability distribution of AoI, and the joint probability distribution of the observed process and its estimation are derived. Pull-based sampling of multiple heterogeneous CTMC-based sources is considered in \cite{akar2024query} where the monitors query the sources according to a Poisson process and the optimum sampling rates for all the sources are obtained that maximize several information freshness metrics including weighted binary freshness.

The analytical models developed in this paper are based on the well-known properties of PH-type distributions which have also been used for information freshness and networked control problems in several existing works. In \cite{akar2023distribution, akar2024age}, the distributions of AoI and peak AoI processes are derived by making use of absorbing Markov chains (AMC) and PH-type distributions. These works can be considered as an alternative to the stochastic hybrid systems (SHS) approach that is widely used to calculate the mean of AoI in \cite{yates2019, chen2023age} and the distribution of AoI \cite{yates2020age, moltafet_etal_tcom22}. In another work \cite{scheuvens2021state}, PH-type distributions are used to find the expected time before a certain number of consecutive packet failures occur in a wireless closed-loop control system.  

\section{Preliminaries} \label{sec:PD}

\subsection{Notation}
Throughout the paper, we use bold lowercase and uppercase characters for a vector, and a matrix, respectively. ${a}_{m}$ denotes the $m$th entry of $\bm{a}$, and $a_{mn}$ denotes the $(m,n)$th element of $\bm{A}$. $\bm{a}^{(-m)}$ refers to the vector $\bm{a}$ by excluding its $m$th entry, and $\bm{A}^{(-m)}$ refers to matrix $\bm{A}$ by excluding both its $m$th row and $m$th column. $\bm{a}^\intercal$ denotes the transpose of vector $\bm{a}$. $\mathrm{e}^{\bm{A}}$ stands for the matrix exponential function $\mathrm{e}^{\bm{A}}=\sum_{k=0}^\infty \dfrac{1}{k!}\bm{A}^k$ \cite{wilcox1967exponential}. $\bm{I}$ denotes the identity matrix. $\bm{1}$ denotes a column vector of ones, $\bm{e}_k$ denotes a column vector of zeros except for the $k$th entry which is one, and $\bm{J}_K$ denotes a square matrix of ones of size $K$.

\subsection{Phase-Type (PH) Distribution}
The following is needed to follow the proposed analytical method for continuous-time absorbing Markov chains (AMC) \cite{kemeny1960finite}. Consider an AMC $Y(t)$, $t\geq 0$, with $K$ transient and $L$ absorbing states. Let the generator of this CTMC be written as,
\begin{align}
 \left[ \begin{array}{c|c}
   \bm A & \bm B \\
   \midrule
   \bm{0} & \bm{0} \\
\end{array}\right], \label{eq:Tmatrix}
\end{align}
where $\bm A_{K \times K}$ and $\bm B_{K \times L}$ are namely the transient sub-generator and absorption sub-generator, and they correspond to the transition rates among the transient states, and from the transient states to the absorbing states, respectively. In this case, we say $Y(t)$ is an AMC characterized with the triple $(\bm A,\bm B,\bm{\beta})$, i.e., $Y(t) \sim AMC(\bm A,\bm B,\bm{\beta})$, where $\bm{\beta} = \{ \bm{\beta}_i \}$ is a ${1 \times K}$ row vector with $\bm{\beta}_i$ denoting the initial probability of being in transient state $i$. 

Upon merging all the absorbing states into one, the distribution of time until absorption, denoted by $T$, is known as the phase-type (PH-type) distribution \cite{neuts81,latouche1999introduction} whose probability density function (PDF) denoted by $f_T(t)$, and cumulative density function (CDF) denoted by $F_T(t)$, respectively, are written in terms of the matrix exponential function of $t\bm A$ as,
\begin{align}
    f_T(t)=-\bm{\beta} \mathrm{e}^{t\bm A}\bm{A}\bm{1}, \quad F_T(t)=1-\bm{\beta} \mathrm{e}^{t\bm A}\bm{1}.
    \label{eq:pdf}
\end{align}
Consequently, the mean and second moment of $T$ can be written as,
\begin{align}
    \mathbb{E}[T]=-\bm{\beta} \bm A^{-1} \bm{1}, \quad
    \mathbb{E}[T^2]=2\bm{\beta} \bm A^{-2} \bm{1}. \label{eq:mom12}
\end{align}
Moreover, the probability of being absorbed in absorbing state $i$ is \cite{kemeny1960finite},
\begin{align}
    p_{i} =-\bm{\beta} \bm{A}^{-1} \bm{B} \bm{e}_{i}. \label{eq:prob}
\end{align} 

The AMC whose generator is given in \eqref{eq:Tmatrix} can be converted to a DTMC with probability transition matrix of the same size as, 
\begin{align}
 \left[ \begin{array}{c|c}
   \bm D & \bm E \\
   \midrule
   \bm{0} & \bm{0} \\ 
\end{array}\right], \label{eq:TmatrixDiscrete1}
\end{align}
at the embedded epochs of state transitions. In particular, $\bm D = \{ d_{ij} \}$ can be written as, 
\begin{align}
    d_{ij}=\begin{cases}
        -\dfrac{a_{ij}}{a_{jj}}, & \text{if } i\neq j, \\ 0, & \text{otherwise.}
    \end{cases} \label{eq:diag}
\end{align}
A basic property about an AMC is that the expected number of visits to a transient state $j$ starting from a transient state $i$ is given by the $(i,j)$th entry of the so-called fundamental matrix $\bm F=(\bm I-\bm D)^{-1}$ \cite{kemeny1960finite}.

\subsection{Multi-Regime Phase-Type (MRPH) Distribution}
In this subsection, we describe the MRPH distribution that is first introduced in the current paper, to the best of our knowledge. For this purpose, we consider a process $Y(t)$, $t \geq 0$ again with $K$ transient and $L$ absorbing states. However, in contrast to the ordinary PH-type distribution, the process $Y(t)$ cannot be characterized with a single generator since the transition rates change with respect to the elapsed time in a piece-wise constant manner. Specifically, there are $M$ finite and distinct boundary values $0=\gamma_1<\gamma_2< \cdots < \gamma_M<\gamma_{M+1}=\infty$ and the behavior of $Y(t)$ is governed by a generator matrix $\bm{Q}_m$, $m=1,\ldots,M$, when $\gamma_m \leq t < \gamma_{m+1}$, i.e., when the AMC is in the {$m$th regime}. We call the distribution of time until absorption, denoted by $T$, a multi-regime PH-type (MRPH) distribution. Note that the ordinary PH-type distribution is a sub-case of the MRPH distribution with a single regime. Similar to the single-regime case, the generator matrices can be expressed for the $m$th regime as, 
\begin{align}
   \bm{Q}_m = \left[ \begin{array}{c|c}
   \bm{A}_m & \bm{B}_m \\
   \midrule
   \bm{0} & \bm{0} \\
\end{array}\right], \label{eq:Tmatrix_MR}
\end{align}
where $\bm{A}_m$ and $\bm{B}_m$ are transition and absorbing sub-generators respectively, in the $m$th regime.
Let us define the probability vector $\bm{g}_Y(t)$ as, 
\begin{align}
    \bm{g}_Y(t)= \begin{bmatrix}
        g_1(t) & g_2(t) & \cdots & g_K(t)
    \end{bmatrix},
\end{align}
where $g_k(t)$ is the probability that the process is in state $k$, $1 \leq k \leq K,$ at time $t$, i.e., $g_k(t)=\mathbb{P}(Y(t)=k)$. One can write the following expression for $\bm{g}_Y(t)$,
\begin{align}
    \bm{g}_Y(t)= 
        \bm{\beta}_m \mathrm{e}^{\bm{A}_m (t-\gamma_m)}, \quad \gamma_m\leq t < \gamma_{m+1},\quad 1\leq m\leq M,  \label{eq:g_yt}
\end{align}
where $\bm{\beta}_m = \bm{g}_Y(\gamma_m)$ corresponds to the initial probability vector of the $m$th regime. For the first regime, the initial probability vector $\bm{\beta}_{1}$ satisfies the condition $\bm{\beta}_{1} \bm{1}=1$, and we can calculate the remaining initial probability vectors from Kolmogorov's forward equation \cite{durrett1999essentials} as, 
\begin{align} \label{eq:kolg_fw}
    \bm{\beta}_{m+1}=\bm{\beta}_m \mathrm{e}^{\bm{A}_m (\gamma_{m+1}-\gamma_m)}, \quad 1\leq m \leq M-1.
\end{align}
Notice that these initial probability vectors ensure the continuity of $\bm{g}_Y(t)$. 
We also observe that the CDF of $T$, denoted by $F_T(t)$, can be written as, 
\begin{align}
    F_T(t)= & \mathbb{P}(T\leq t) 
          = 1-\sum_{k=1}^K g_k(t) = 1-\bm{g}_Y(t) \bm{1}, \\
          =& 1-\bm{\beta}_m \mathrm{e}^{\bm{A}_m (t-\gamma_m)} \bm{1}, \quad \gamma_m\leq t < \gamma_{m+1},\quad 1 \leq m\leq M.  
\end{align}
Consequently, the PDF of $T$, denoted by $f_T(t)$, can be obtained by differentiating the above expression,
\begin{align}
    f_T(t)= -\bm{\beta}_m \mathrm{e}^{\bm{A}_m (t-\gamma_m)} \bm{A}_m \bm{1}, \quad \gamma_m\leq t < \gamma_{m+1},\quad 1 \leq m\leq M, \label{eq:pdf_mr}
\end{align}
which yields the following expression for $\mathbb{E}[T]$,
\begin{align}
    \mathbb{E}[T]&= \sum_{m=1}^{M-1} \int_{\gamma_m}^{\gamma_{m+1}} t f_T(t)  \dd{t} + \int_{\gamma_M}^{\infty} t f_T(t)  \dd{t},  \\
    &=\sum_{m=1}^{M-1} \underbrace{\int_{\gamma_m}^{\gamma_{m+1}} -t\bm{\beta_m} \mathrm{e}^{\bm{A}_m(t-\gamma_m)} \bm{A}_m\bm{1}  \dd{t}}_{I_{1}(\bm{\beta}_m,\bm{A}_m,\gamma_{m},\gamma_{m+1}) }+\underbrace{\int_{\gamma_M}^{\infty} -t\bm{\beta}_m \mathrm{e}^{\bm{A}_m(t-\gamma_M)} \bm{A}_m\bm{1} \dd{t} }_{I_{1,\infty}(\bm{\beta}_m,\bm{A}_m,\gamma_{M})}, \label{eq:first_moment_ns}
\end{align}
where the expressions for the integrals denoted by $I_{1}(\bm{\beta},\bm{A},\gamma_{l},\gamma_{u})$ and $I_{1,\infty}(\bm{\beta},\bm{A},\gamma)$, respectively, can be written as,
\begin{align}
   I_{1}(\bm{\beta},\bm{A},\gamma_{l},\gamma_{u}) &=-\bm{\beta} \mathrm{e}^{\bm{A}(\gamma_{u}-\gamma_l)} (\gamma_{u} \bm{A}-\bm{I} ) \bm{A}^{-1} \bm{1} +\bm{\beta} (\gamma_{l} \bm{A}-\bm{I}) \bm{A}^{-1}\bm{1}, \\ 
   I_{1,\infty}(\bm{\beta},\bm{A},\gamma) &= \bm{\beta}(\gamma \bm{A}-\bm{I})\bm{A}^{-1}\bm{1}.
\end{align}
Similarly, the second moment of the time before absorption can be expressed as, 
\begin{align}
    \mathbb{E}[T^2]&=  \sum_{m=1}^{M-1} \int_{\gamma_m}^{\gamma_{m+1}} t^2 f_T(t)  \dd{t} + \int_{\gamma_M}^{\infty} t^2 f_T(t)  \dd{t},   \\
    &=\sum_{m=1}^{M-1} \underbrace{\int_{\gamma_m}^{\gamma_{m+1}} -t^2\bm{\beta}_m \mathrm{e}^{\bm{A}_m(t-\gamma_m)} \bm{A}_m\bm{1}  \dd{t}}_{I_{2}(\bm{\beta}_m,\bm{A}_m,\gamma_{m},\gamma_{m+1})}+\underbrace{\int_{\gamma_M}^{\infty} -t^2\bm{\beta}_m \mathrm{e}^{\bm{A}_m(t-\gamma_M)} \bm{A}_m\bm{1} \dd{t} }_{I_{2,\infty}(\bm{\beta}_m,\bm{A}_m,\gamma_{M})}, \label{eq:second_moment_ns}
\end{align}
where the expressions for the integrals denoted by $I_{2}(\bm{\beta},\bm{A},\gamma_{l},\gamma_{u})$ and $I_{2,\infty}(\bm{\beta},\bm{A},\gamma)$, respectively, can be written as,
\begin{align}
   I_{2}(\bm{\beta},\bm{A},\gamma_{l},\gamma_{u}) =&-\bm{\beta} \mathrm{e}^{\bm{A_m}(\gamma_{m+1}-\gamma_m)} (\gamma_{m+1}^2 \bm{A_m}^2-2\gamma_{m+1}\bm{A_m}+2\bm{I} ) \bm{A_m}^{-2} \bm{1} \nonumber\\
   &-\bm{\beta_m} ( \gamma_{m}^2 \bm{A_m}^2-2\gamma_{m}\bm{A_m}+2\bm{I} ) \bm{A_m}^{-2}\bm{1},  \\ 
   I_{2,\infty}(\bm{\beta},\bm{A},\gamma) =& - \bm{\beta}(\gamma^2 \bm{A}^2-2\gamma\bm{A}+2\bm{I})\bm{A}^{-2}\bm{1}.
\end{align}

We are also interested in obtaining the absorption probabilities for which purpose we first define an absorption rate vector $\bm{\nu}(t)$ as,
\begin{align}
    \bm{\nu}(t)= \begin{bmatrix}
        \nu_1(t) & \nu_2(t) & \cdots & \nu_L(t)
    \end{bmatrix},
\end{align}
where ${\nu}_l(t) \delta t$ is the probability that $Y(t)$ is absorbed in the time interval $(t, t + \delta t)$ into absorbing state $l$ as $\delta t \rightarrow 0$. Recall that for regime $m$, the $(k,l)$th element of the matrix $\bm{B}_m$ corresponds to the rate at which $Y(t)$ is absorbed into absorbing state $l$ when the process is in state $k$. Therefore, by multiplying the probability vector $\bm{g}_Y(t)$ with the $l$th column of $\bm{B}_m$, which we denote by $\bm{b}_{m,*l}$, we can write 
${\nu}_l(t)$ as,
\begin{align}
    {\nu}_{l}(t)=
        \bm{\beta}_m \mathrm{e}^{\bm{A}_m(t-\gamma_m)}\bm{b}_{m,*l}, \quad  \gamma_m\leq t < \gamma_{m+1},\ 1 \leq m\leq M.
\end{align}
Note that the summation $f_T(t) = \sum_{l=1}^L\nu_{l}(t)$ by definition. 

Finally, we can calculate the probability that the process $Y(t)$ is absorbed into the $l$th absorption state, denoted by $p_l$, as follows,
\begin{align} \label{eq:probij}
    p_l&= \int_{0}^{\infty} \nu_{l}(t)\dd{t}=\sum_{m=1}^{M-1} \underbrace{\int_{\gamma_m}^{\gamma_{m+1}} \bm{\beta}_m \mathrm{e}^{\bm{A}_m(\gamma_{m+1}-\gamma_m)}\bm{b}_{m,*l}  \dd{t}}_{I_{0}(\bm{\beta}_m,\bm{A}_m,\bm{b}_{m,*l},\gamma_m,\gamma_{m+1})}+\underbrace{\int_{\gamma_M}^{\infty} \bm{\beta}_M \mathrm{e}^{\bm{A}_M(t-\gamma_M)}\bm{b}_{M,*l} \dd{t} }_{I_{0,\infty}(\bm{\beta}_M,\bm{A}_M,\bm{b}_{M,*l},\gamma_M)},
\end{align}
where the values of the integrals denoted by $I_{0}(\bm{\beta},\bm{A},\bm{b},\gamma_l,\gamma_{u})$ and $I_{0,\infty}(\bm{\beta},\bm{A},\bm{b},\gamma)$, respectively, can be written as,
\begin{align}
   I_{0}(\bm{\beta},\bm{A},\bm{b},\gamma_l,\gamma_{u}) &=\bm{\beta} \left(\mathrm{e}^{\bm{A}(\gamma_{u}-\gamma_l)}-\bm{I}\right)\bm{A}^{-1} \bm{b}, \\ 
   I_{0,\infty}(\bm{\beta},\bm{A},\bm{b},\gamma) &= - \bm{\beta} \bm{A}^{-1} \bm{b} .
\end{align}

\section{System Model} \label{sec:sys}
We consider a finite, irreducible (and hence ergodic) CTMC process $X(t) \in \mathcal{N}=\{1,2,\ldots,N\}$, with generator matrix $\bm{Q} = \{ q_{ij} \}$, with non-negative off-diagonal elements  corresponding to the transition rates between states, and with negative diagonal elements $q_{ii} = - \sigma_i$, i.e., holding rate at state $i$, 
\begin{align}
    Q=\begin{bmatrix}
        -\sigma_1 & q_{12} & \dots & q_{1N} \\
        q_{21} & -\sigma_2 & & \vdots \\
        \vdots & & \ddots &  \\
        q_{N1} & \dots & & -\sigma_N
    \end{bmatrix},  \label{eq:QMat}
\end{align} 
and the holding time (or sojourn time) at state $i$, denoted by $H_i$, is exponentially distributed with parameter $\sigma_i$, i.e., $H_i\sim \text{Exp}(\sigma_i)$. We refer the reader to \cite{durrett1999essentials, norris1998markov} for a detailed treatment of CTMCs and their properties. At the end of the holding time, a state transition to another state $j \neq i$ occurs with probability $\rho_{ij}=\dfrac{q_{ij}}{\sigma_i}$.

The source's sensor observes the process $X(t)$ through sampling and generates packets carrying sample values with the generate-at-will (GAW) principle. Each generated packet is transmitted towards the remote monitor via a communications channel where the transmission time (or service time), denoted by $E$, is assumed to be exponentially distributed with parameter $\mu$, i.e., $E \sim \text{Exp}(\mu)$.  The estimation law is $\hat{X}(t) = X(t')$ where $t'$ is the timestamp of the latest status update packet received before time $t$, which is also widely used in existing remote estimation studies \cite{maatouk2020}. We also assume that the source is immediately acknowledged when the transmission is completed with instantaneous feedback from the monitor to the source. The source only starts transmitting the state it observes and immediately preempts an ongoing transmission if the observed state changes before the transmission is completed.

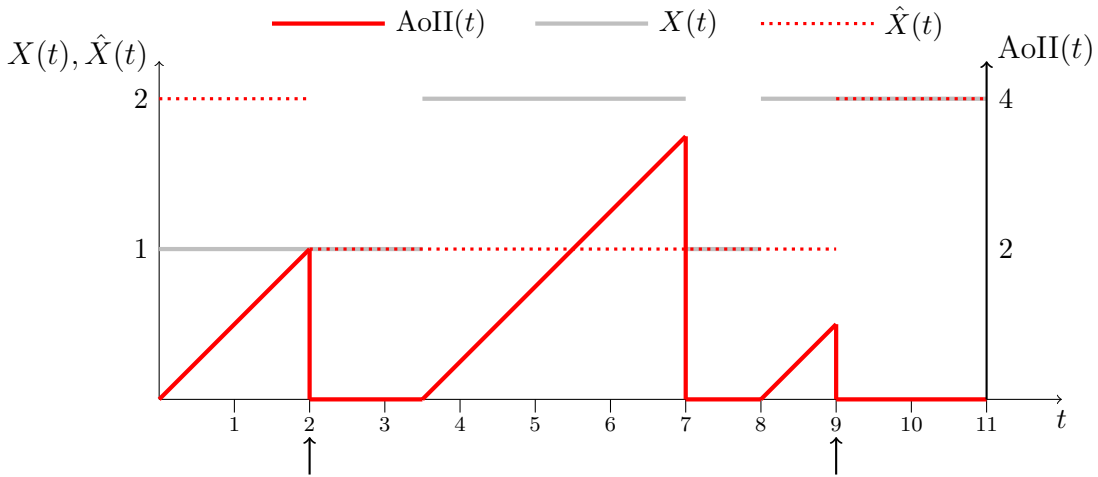
\begin{figure}[tb]
    \centering
    \begin{tikzpicture}[scale=0.50]
    \draw[->] (3,8) -- (27,8);
    \filldraw (27,8) node[anchor=north] {\small{$t$}};
    \draw[->] (3,8) -- (3,17);
    \filldraw (3,18) node[anchor=north east] {{$X(t),\hat{X}(t)$}};
    \draw[-] (5,8) -- (5,7.65);
    \draw[-] (7,8) -- (7,7.65);
    \draw[-] (9,8) -- (9,7.65);
    \draw[-] (11,8) -- (11,7.65);
    \draw[-] (13,8) -- (13,7.65);
    \draw[-] (15,8) -- (15,7.65);
    \draw[-] (17,8) -- (17,7.65);
    \draw[-] (19,8) -- (19,7.65);
    \draw[-] (21,8) -- (21,7.65);
    \draw[-] (23,8) -- (23,7.65);
    \draw[-] (25,8) -- (25,7.65);
    
    \filldraw (7,7.8) node[anchor=north] {\scriptsize{2}};
    \filldraw (11,7.8) node[anchor=north] {\scriptsize{4}};
    \filldraw (15,7.8) node[anchor=north] {\scriptsize{6}};
    \filldraw (19,7.8) node[anchor=north] {\scriptsize{8}};
    \filldraw (23,7.8) node[anchor=north] {\scriptsize{10}};
    \filldraw (25,7.8) node[anchor=north] {\scriptsize{11}};
    \filldraw (5,7.8) node[anchor=north] {\scriptsize{1}};
    \filldraw (9,7.8) node[anchor=north] {\scriptsize{3}};
    \filldraw (13,7.8) node[anchor=north] {\scriptsize{5}};
    \filldraw (17,7.8) node[anchor=north] {\scriptsize{7}};
    \filldraw (21,7.8) node[anchor=north] {\scriptsize{9}};
    
    \draw[red, ultra thick] (6,18) -- (9,18);
    \filldraw (9,18) node[anchor=west] {\small{${\text{AoII}(t)}$}};
    \draw[lightgray, ultra thick] (13,18) -- (16,18);
    \filldraw (16,18) node[anchor=west] {\small{$X(t)$}};
    \draw[dotted, red, very thick] (19,18) -- (22,18);
    \filldraw (22,18) node[anchor=west] {\small{$\hat{X}(t)$}};
    
    \draw[->,black,thick] (7,6) -- (7,7);
    \draw[->,black,thick] (21,6) -- (21,7);
    
    \filldraw (3,12) node[anchor=east] {\small{1}};
    \draw[lightgray,ultra thick] (3,12) -- (10,12);
    \filldraw (3,16) node[anchor=east] {\small{2}};
    \draw[dotted,red,very thick] (3,16) -- (7,16);
    
    \draw[lightgray,ultra thick] (10,16) -- (17,16);
    \draw[lightgray,ultra thick] (17,12) -- (19,12);
    \draw[lightgray,ultra thick] (19,16) -- (25,16);
    \draw[dotted,red,very thick] (7,12) -- (21,12);
    \draw[dotted,red,very thick] (21,16) -- (25,16);
    
    \draw[red,ultra thick] (3,8) -- (7,12);
    \draw[red,ultra thick] (7,12) -- (7,8);
    \draw[red,ultra thick] (7,8) -- (10,8);
    \draw[red,ultra thick] (10,8) -- (17,15);
    \draw[red,ultra thick] (17,15) -- (17,8);
    \draw[red,ultra thick] (17,8) -- (19,8);
    \draw[red,ultra thick] (19,8) -- (21,10);
    \draw[red,ultra thick] (21,10) -- (21,8);
    \draw[red,ultra thick] (21,8) -- (25,8);

    \filldraw (26,12) node[anchor=east] {\small{2}};
    \filldraw (26,16) node[anchor=east] { \small{4}};
    \draw[black,thick,->] (25,8) -- (25,17);
    \filldraw (25,18) node[anchor=north west] {$\text{AoII}(t)$};
    \end{tikzpicture}
    \caption{Sample path of $X(t)$ (light grey lines), $\hat{X}(t)$ (red dotted lines), and $\text{AoII}(t)$ (thick red solid curve) for an example scenario when $N=2$ and $\text{AoII}(0)=0$. The arrows at $t=2$ and $t=9$ represent the reception epochs of status update packets at the monitor. Notice that $\text{AoII}(t)$ drops to zero at $t=7$ without a packet reception, as the process $X(t)$ returns on its own to the current estimate at the monitor $\hat{X}(t)$.}
    \label{fig:SamplePath}
\end{figure}

The mismatch between $X(t)$ and $\hat{X}(t)$ is measured with the AoII metric which is generically defined in \cite{maatouk2020} as \( \text{AoII}(t) = f_{\text{time}}(t) \times f_{\text{mismatch}} (X(t), \hat{X}(t) ), \) where $f_{\text{time}}(t)$ is a time penalty function that increases while the mismatch stays, and 
$f_{\text{mismatch}}(X(t),\hat{X}(t))$ is a mismatch penalty depending on the instantaneous values of the pair $(X(t),\hat{X}(t))$. In this paper, we focus on the linear time penalty function, and the indicator function for mismatch penalty. This choice gives rise to the following simplified definition of AoII used in the current paper,
\begin{align}
    \text{AoII}(t)=t-\max(\zeta \; | \; X(\zeta)=\hat{X}(\zeta),\; \zeta \leq t). \label{eq:aoii}
\end{align}
However, it is possible to adopt more general AoII penalty functions in our framework since the  distribution of the mismatch time can also be obtained with the proposed method. A sample path of $X(t)$, $\hat{X}(t)$ and $\text{AoII}(t)$ is illustrated in Fig.~\ref{fig:SamplePath} for an example CTMC with two states.

\section{Optimum Transmission Policy} \label{sec:esat}
In this section, we consider a transmission policy for which the sensor initiates a transmission if $\text{AoII}(t)$ exceeds a threshold value $\tau_{ji}$ when $X(t)=i, \ \hat{X}(t)=j$. First, we formulate the optimization problem as a CSMDP by defining a DTMC at the embedded epochs of synchronization. Then, we analyze the performance for given thresholds $\tau_{ji}$, and obtain the parameters of the underlying CSMDP. Finally, we conclude the section by showing that this transmission policy is optimum for a general finite-state irreducible CTMC source.

\subsection{CSMDP Formulation}
In this subsection, we study the process $\text{AoII}(t)$ in cycles where a cycle starts and ends at two successive synchronization points of $X(t)$ and $\hat{X}(t)$. Then, we define an embedded DTMC whose states are the corresponding synchronization values. Finally, we formulate the optimization problem as a CSMDP for which purpose we first need the following definition. 

\begin{definition}
    A time point $t_0$ is a synchronization point (SP) at synchronization value $S_i$ when
    $X(t_0)=\hat{X}(t_0)=i$, $X(t_0^-) \neq \hat{X}(t_0^-)$, i.e., when $X(t)$ and $\hat{X}(t)$ are just synchronized at state $i$ at time $t_0$. 
\end{definition}

The interval between the SP $t_0$ at $S_i$ and the next SP is called cycle-$i$, i.e., a cycle that starts with the synchronization value $S_i$. In this notation, $S_i$ is used to denote a state of the embedded DTMC, whereas $i$ is used for indexing the states of the original CTMC.  We now consider the DTMC embedded at the SPs with the states being the synchronization values $S_i$, $i \in \mathcal{N}$. The action taken at state $S_j$ at the corresponding SP is a threshold vector $\bm{\tau}_j=\{\tau_{ji}\}$ for $j\in\mathcal{N}$, $i\in\mathcal{N}\backslash j$, and $\bm{\tau}_j \in{\rm I\!R}_+^{N-1}$, and we denote the collection of all thresholds by an $N-1 \times N$ matrix $\bm{T}=\left[\bm{\tau}_1, \dots, \bm{\tau}_N \right]$. The duration of the cycle corresponding to $S_j$, $D_j(\bm{\tau}_j)$, is a function of the threshold vector whose expected value is denoted by $d_j(\bm{\tau}_j)$. In our problem, we consider two types of costs which also depend on the threshold values. Denoting the area under the $\text{AoII}(t)$ curve in cycle-$j$ by $A_j(\bm{\tau}_j)$ whose expected value gives the AoII-cost which is denoted by $a_j(\bm{\tau}_j)$. Similarly, denoting the number of transmissions attempted in cycle-$j$ by $C_j(\bm{\tau}_j)$, its expected value, denoted by $c_j(\bm{\tau}_j)$, is termed as the sampling cost in our formulation. Finally, $b$ denotes the sampling budget which bounds the number of transmission attempts per unit time.

While the transition from state $S_j$ (at the current SP) to $S_i$ (at the next SP) for $j \neq i$ is incurred with a timely reception of a sample, a self-transition from state $S_j$ to itself can also occur when $X(t)$ returns to its initial value $j$ without a new sampling operation. Let $p_{ji}(\bm{\tau}_j)$ and $\pi_j$ denote the transition probability from synchronization value $S_j$ to $S_i$ for threshold vector $\bm{\tau}_j$, and the steady-state probability of being in state $S_j$ at an SP, respectively. The steady-state vector $\boldsymbol{\pi} = \{ \pi_j\}$ satisfies,
\begin{align}
    \bm{\pi}&=\bm{\pi} \bm{P}, \quad \bm{P} \bm{1}=\bm{1}, 
\end{align}
where $\bm P=\{ p_{ji}(\bm{\tau}_j) \}$, and   $\bm{\pi}$ can be obtained by solving a linear matrix equation,
\begin{align}
    \bm{\pi } =\bm{1}^{\intercal}\left(\bm{P}+\bm{1}\bm{1}^{\intercal}-\bm{I}\right)^{-1}. \label{eq:oneonetranspose}
\end{align}
Note that both $\bm{\pi }$ and $\bm P$ defined above depend on the matrix $\bm T$.

The mean AoII, denoted by $\text{MAoII}$, where AoII stands for the steady-state random variable associated with the marginal of the stationary stochastic process $\text{AoII}(t)$, is also the following time average due to the ergodicity of the CTMC $X(t)$,
\begin{align}
    \text{MAoII}= \lim_{T\to\infty} \frac{1}{T} \int_{0}^T \text{AoII}(t) \dd{t}.
\end{align}
Let $N(t)$ denote the number of samples taken (or transmissions attempted) in the interval $[0,t]$. We define the average sampling rate $R$ as,
\begin{align}
    R=\lim_{T\to\infty} \frac{1}{T} \int_{0}^T N(t) \dd{t}.
\end{align}
Then, the optimization problem of interest is cast as,  
\begin{mini}
	{\phi\in\Phi}{\text{MAoII}^{\phi}} 
	{\label{Opt1}}
    {}
	\addConstraint{ R^{\phi} }{\leq b} 
\end{mini}
over all sampling policies $\phi\in\Phi$, with $\text{MAoII}^{\phi}$ denoting the average AoII and $R^{\phi}$ denoting the average sampling rate, obtained when policy $\phi$ is imposed. Here, $\phi$ refers to a deterministic mapping from each state $S_j$, $j \in \mathcal{N}$ to a set of AoII thresholds that will be used to initiate transmission operations during cycle-$j$. If $s_n\in \mathcal{N}$ is the state at the $n$th decision epoch for policy $\phi$, then the quantities $\text{MAoII}^{\phi}$ and $R^\phi$ can be expressed as,
\begin{align}
   \text{MAoII}^{\phi}&=\lim_{L\to\infty} \dfrac{ \mathbb{E} \left[\sum_{n=1}^L A_{s_n}(\bm{\tau}_{s_n}^{\phi}) \right] }{ \mathbb{E} \left[\sum_{n=1}^L D_{s_n}(\bm{\tau}_{s_n}^{\phi})\right] }, \label{eq:MaoII} \\
    R^\phi&= \lim_{L\to\infty} \dfrac{ \mathbb{E}\left[\sum_{n=1}^L C_{s_n}(\bm{\tau}_{s_n}^{\phi}) \right] }{ \mathbb{E} \left[\sum_{n=1}^L D_{s_n}(\bm{\tau}_{s_n}^{\phi})\right] }, \label{eq:R}
\end{align}
where $\bm{\tau}_{s_n}^{\phi}$ is the collection of thresholds for state $s_n$ related to the policy $\phi$. These expressions can also be written in terms of the policy thresholds as,
\begin{align}
   \mbox{MAoII}^{\phi} & =\dfrac{\sum_{j=1}^{N}\pi_j a_j(\bm{\tau}_j^{\phi})}{\sum_{j=1}^{N} \pi_j d_j(\bm{\tau}_j^{\phi})}, \label{eq:age_app} \\
     R^{\phi}& =\dfrac{\sum_{j=1}^{N}\pi_j c_j(\bm{\tau}_j^{\phi})}{\sum_{j=1}^{N} \pi_j d_j(\bm{\tau}_j^{\phi})}. \label{eq:rate_app}
\end{align}

The Lagrangian method is used to solve the constrained SMDP by converting it to an unconstrained SMDP with a Lagrangian multiplier $\lambda$ and iteratively solving it for different values of $\lambda$ \cite{makowski1986}. In this case, we have the unconstrained optimization problem,
\begin{mini}
	{\phi}{J^{\phi}(\lambda)} 
	{\label{OptUc}}
    {}
\end{mini}
where $J^{\phi}(\lambda)=\text{MAoII}^{\phi}+\lambda R^{\phi}$ can be expressed as,
\begin{align}
    J^{\phi}(\lambda) & =\lim_{L\to\infty} \dfrac{ \mathbb{E}^\phi \left[\sum_{n=1}^L A_{s_n}(\bm{\tau}_{s_n}^{\phi})+\lambda C_{s_n}(\bm{\tau}_{s_n}^{\phi}) \right] }{ \mathbb{E}^\phi \left[\sum_{n=1}^L D_{s_n}(\bm{\tau}_{s_n}^{\phi})\right] },
\end{align}
which can also alternatively be written in terms of the policy thresholds as,
\begin{align}
    J^{\phi}(\lambda) & =\dfrac{\sum_{j=1}^{N}\pi_j \left[a_j(\bm{\tau}_j^{\phi})+\lambda c_j(\bm{\tau}_j^{\phi})\right]}{\sum_{j=1}^{N} \pi_j d_j(\bm{\tau}_j^{\phi})}. 
\end{align}
Finally, we define $\text{MAoII}(\lambda)$ and $R(\lambda)$, respectively, as the objective and constraint functions, obtained from the minimization of \eqref{OptUc} for given parameter $\lambda$.
In the same work \cite{makowski1986}, it is shown that there exists a Lagrangian coefficient $\lambda^*$ such that the optimum policy $\phi^*$ obtained from the unconstrained problem is also optimum for the constrained problem either when (i) the constrained problem attains $R^{\phi^*}=b$, or (ii) $\lambda=0$ and $R^{\phi^*}\leq b$.

For the unconstrained problem in \eqref{OptUc}, we propose to use the policy iteration algorithm given in \cite{white1993mdp}. First, we write the optimality equations for the unconstrained problem based on \cite{ibe2013markov} as,
\begin{align}
    V_j&=\min_{ \bm{\tau}_j \in{\rm I\!R}_+^{N-1} } V_j(\bm{\tau}_j)=\min_{\bm{\tau}_j \in{\rm I\!R}_+^{N-1}}\left\{a_j(\bm{\tau}_j)+\lambda c_j{(\bm{\tau}_j)}-\eta d_{j}(\bm{\tau}_j)+\sum_{j =1 }^N p_{ji}(\bm{\tau}_j) V_{j}(\bm{\tau}_j)\right\}, \label{eq:opt_gen}
\end{align}
where $V_j$ denotes the optimum average cost starting from initial state $S_j$ and $V_j(\bm{\tau}_j )$ is the average cost attained when the action $\bm{\tau}_j \in{\rm I\!R}_+^{N-1}$ is applied at initial state $S_j$ for the first cycle-$j$ but then the optimum policy is applied. On the other hand, $\eta$ corresponds to the long-term average cost. The policy iteration algorithm (see \cite{white1993mdp} for details) is presented in Algorithm~\ref{alg:cap_ij}.

\begin{algorithm}
    \caption{Policy iteration algorithm for the unconstrained CSMDP}\label{alg:cap_ij}
    \begin{algorithmic}
        \State \textbf{Initialize:} $\bm{\tau}_{j}=\bm{0}$ for $j\in \mathcal{N}$.
        \State \textbf{Step 1: (CSMDP model)}  Obtain the values $a_j(\bm{\tau}_j)$, $c_j(\bm{\tau}_j)$, $d_j(\bm{\tau}_j)$ and $p_{ji}(\tau_j)$ for given vector of threshold vector $\bm{\tau}_j$ for each $j$. 
        \State \textbf{Step 2: (value determination)}: Obtain the long-time average cost $\eta$ and the relative values $V_j$ for $1 \leq j <N$ by fixing $v_N=0$ and solving the following $N$ optimality equations 
        \begin{align}
            V_j =a_j(\bm{\tau}_j)+\lambda c_j(\bm{\tau}_j) -\eta d_j(\bm{\tau}_j)+\sum_{i=1}^N p_{ji}(\bm{\tau}_j)V_j.
        \end{align}
        \State \textbf{Step 3: (policy improvement)}: For each $j$, set $\bm{\tau}_j$ to 
        \begin{align}
            \underset{\bm{\tau}_j   \in{\rm I\!R}_+^{N-1}}{\arg \min}    \dfrac{a_{j}(\bm{\tau}_j)+\lambda c_j(\bm{\tau}_j)}{d_j(\bm{\tau}_j)}+\dfrac{\sum_{j\neq i} (v_{j}-v_i) p_{ji}(\bm{\tau}_j) }{d_j(\bm{\tau}_j)}. \label{eq:pol}
        \end{align}
        \State \textbf{Step 4: (stopping rule)} If $|\eta^{(n)}-\eta^{(n-1)}|\leq \epsilon_\eta$ then stop. Otherwise go to Step 1.  Here, $\eta^{(n)}$ denotes the long-time average cost obtained at iteration $n$.
    \end{algorithmic}
\end{algorithm}

The policy iteration algorithm iteratively evaluates and improves policies until convergence \cite{white1993mdp} to solve the unconstrained problem. Additionally, it is shown in \cite{sennott1993constrained} that the constraint function ($R(\lambda)$ in our case) obtained by minimizing the Lagrangian problem in \eqref{OptUc} is a monotonically decreasing function of the Lagrangian coefficient $\lambda$. Thus, the constraint set $\{\lambda|R(\lambda)\leq b, 0 \leq \lambda\}$ is a convex set. Then, the optimum thresholds that minimize the constrained problem are obtained as follows. First, the policy iteration algorithm is run for $\lambda=0$. If the average sampling rate (which can be calculated by \eqref{eq:R} for the obtained policy), is less than the budget $b$, then we have the optimum solution. Otherwise, the optimum Lagrange multiplier $\lambda^*$ with the corresponding thresholds satisfying the constraint on the boundary is found by line search.

\subsection{Estimation and State Aware Transmission (ESAT) Policy}
In this subsection, we propose a method to find the CSMDP parameters $d_j(\bm{\tau}_j)$, $a_j(\bm{\tau}_j)$, $c_j(\bm{\tau}_j)$ and $p_{ji}(\bm{\tau}_j)$ for cycle-$j$ given the threshold vector $\bm{\tau}_j$ to be used in cycle-$j$. Then, the policy iteration algorithm given in Algorithm~\ref{alg:cap_ij} can be used to solve the unconstrained CSMDP given in \eqref{OptUc} which is an intermediate step in the solution of the CSMDP stated in \eqref{Opt1}, with the method detailed in the previous subsection.

\begin{figure}[t]
    \centering
   \includegraphics[width=0.85\columnwidth]{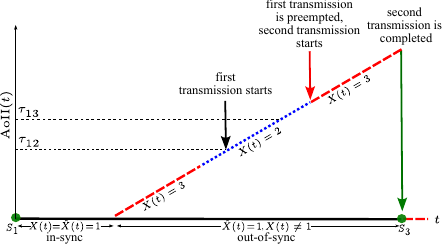} 
    \caption{A sample path for $X(t)$, $\hat{X}(t)$ and $\text{AoII}(t)$ for an example scenario. Green circles denote the synchronization points.}
    \label{fig:trans_ij}
\end{figure}

Fig.~\ref{fig:trans_ij} illustrates a sample cycle-$1$ between the synchronization values $S_1$ and $S_3$ for an example scenario with $N=3$ for which $\tau_{12}<\tau_{13}$. As long as the system is in-sync at value $S_1$, AoII$(t)$ is zero. The synchronization is broken when $X(t)$ transitions to state $3$ and the system enters the out-of-sync phase during which AoII$(t)$ increases linearly with time. When $X(t)=3$, $\text{AoII}(t)$ does not exceed the threshold, and transmission is not initiated. However, after $X(t)$ transitions from state $3$ to state $2$, $\text{AoII}(t)$ hits the threshold $\tau_{12}$ at which instant the first transmission is initiated. However, this transmission is preempted when $X(t)$ transitions back to state $3$. At this time point, the threshold $\tau_{13}$ has already been exceeded which triggers a second transmission. When this transmission is complete, the new synchronization value $S_3$ is reached. Notice that the total number of attempted transmissions is two whereas there is only one completed transmission in this example. 

For cycle-$j$, we define an AMC process $Y_{j}(t)$, $t \geq 0$, which will start evolving when synchronization is just broken and this process will evolve until absorption occurs when the source and monitor synchronize again.
Therefore, the time until absorption of this AMC process, denoted by $T_j$, overlaps with the out-of-sync duration in cycle-$j$.
Thus, we can write,
\begin{align}
    D_j(\bm{\tau}_j)=H_j+T_j, \quad
    A_j(\bm{\tau}_j)=\dfrac{T_j^2}{2}. \label{eq:ADij}
\end{align}
The process $Y_{j}(t)$ has $N-1$ transient states which represent the value of $X(t)=i$, $i\in \mathcal{N}\backslash j$, and $N$ absorbing states, which are the synchronization values $S_i$, $i\in \mathcal{N}$. During cycle-$j$, absorption to $S_j$ can happen anytime when $X(t)$ transitions back to the estimated state $j$. However, absorption to other states can take place when the corresponding threshold is exceeded followed by a completed transmission. The transition rates of the AMC process $Y_{j}(t)$ are summarized in Table~\ref{tab:TablensPD}. Therefore, the process $Y_j(t)$ can be defined through a MRPH distribution with $M\leq N$ regimes with each regime $m$ starting from the boundary value $\gamma_{jm}$ which are to be obtained through the threshold vector as follows. We first define a vector $\bm{\gamma}_{j}=\{\gamma_{jm}\}$ corresponding to the regime boundaries. This vector $\bm{\gamma}_{j}$ consists of all distinct elements of $\bm{\tau}_j$, $0$, and $\infty$ and it is sorted as,
\begin{align}
    \bm{\gamma}_{j}=\left\{0,\min(\bm{\tau}_j|\tau_{ji}>\gamma_{j1}), \min(\bm{\tau}_j|\tau_{ji}>\gamma_{j2}), \dots,\max(\bm{\tau}_j|\tau_{ji}>\gamma_{j \, M-1}),\infty \right\},
\end{align}
where $M\leq N$ is the size of the vector $\bm{\gamma}_{j}$, with equality if all elements of $\bm{\tau}_i$ are distinct, and non-zero. Additionally, we define the set $\mathcal{V}_{jm}$ that consists of indices of all thresholds $\tau_{ji}$ such that they are less than or equal to $\gamma_{jm}$. 

\begin{table}[t]
    \caption{Non-zero transition rates for the absorbing CTMC process $Y_j(t)$ for ESAT.}
    \vspace*{0.2cm}
    \centering
    \begin{tabular}{|c|c|c|} 
    \hline
    \multicolumn{3}{|c|}{Transition rates from state $i$ for $Y_j(t)$} \\
    \hline
    To & Condition & Value \\ 
    \hline
    $i'$ & $ i' \neq j $ & $q_{ii'}$ \\
    \hline
    $S_j$ & $ \forall i$ & $q_{ij}$ \\
    \hline
    $S_i$ & $i\neq j$, $t\geq\tau_{ji}$ & $\mu$ \\
    \hline
    \end{tabular}
    \label{tab:TablensPD}
\end{table}

Similar to Section~\ref{sec:PD}, we define a vector $\bm{g}_{Y_j}(t)=\{g_{ji}\}$, $i\in \mathcal{N}\backslash j$, where $g_{ji}(t)$ is the probability that the process is in state $i$ at time $t$, and it is equal to $g_{ji}(t)=\mathbb{P}(Y_{j}(t)=i)$. From \eqref{eq:g_yt}, it can be expressed as,
\begin{align}
    \bm{g}_{Y_j}(t)= 
        \bm{\beta}_{jm} \mathrm{e}^{\bm{A}_{jm} (t-\gamma_{jm})}, \quad \gamma_{jm}\leq t < \gamma_{j \, m+1},\quad 1\leq m\leq M, 
\end{align}
where $\bm{\beta}_{jm}$ is the initial probability of the $m$th regime, and $\bm{A}_{jm}$ and $\bm{B}_{jm}$ are the transition and absorption sub-generator matrices for the $m$th regime, respectively. Moreover, the PDF of $T_j$ can be expressed as in \eqref{eq:pdf_mr},
\begin{align}
    f_{T_j}(t)=
        -\bm{\beta}_{jm} \mathrm{e}^{\bm{A}_{jm} (t-\gamma_{jm})}\bm{A}_{jm}\bm{1}, \quad \gamma_{jm}\leq t < \gamma_{j \, m+1},\quad 1\leq m\leq M.
\end{align}

For cycle-$j$, the initial probability vector of the first regime can be written as, 
\begin{align}
    \bm{\beta}_{j1}=\dfrac{1}{\sigma_j}\bm{q}^{(-j)}_{j*}=\begin{bmatrix}
        \dfrac{q_{j1}}{\sigma_j} & \dfrac{q_{j2}}{\sigma_j} & \dots & \dfrac{q_{j \, j-1}}{\sigma_j} & \dfrac{q_{j \, j+1}}{\sigma_j} & \dots & \dfrac{q_{jN}}{\sigma_j}
    \end{bmatrix}, \label{eq:beta1}
\end{align}
where $\bm{q}^{(-j)}_{j*}$ is the $j$th row of $\bm{Q}$ with the $j$th element removed recalling that the operation $(\cdot)^{(-i)}$ was defined in Section~\ref{sec:PD} as an exclusion operation for vectors and matrices.  The remaining initial probabilities can be obtained from \eqref{eq:kolg_fw} as,
\begin{align}
    \bm{\beta}_{j \, m+1}=\bm{\beta}_{jm} \mathrm{e}^{\bm{A}_{jm} (\gamma_{j \, m+1}-\gamma_{jm})}, \quad 1\leq m \leq M-1. \label{eq:betam}
\end{align}

The next step is to find the transition and absorption sub-generator matrices, namely, $\bm{A}_{jm}$ and $\bm{B}_{jm}$, from Table~\ref{tab:TablensPD} for cycle-$j$, as 
\begin{align}
    \bm{A}_{jm}=\begin{bmatrix}
        \bm{a}_{jm,1*} \\
        \bm{a}_{jm,2*} \\
        \vdots \\
        \bm{a}_{jm,j-1\,*} \\
        \bm{a}_{jm,j+1\,*} \\
        \vdots \\
        \bm{a}_{jm,N*}
    \end{bmatrix}, \quad
    \bm{B}_{jm}=\begin{bmatrix}
        \bm{b}_{jm,1*} \\
        \bm{b}_{jm,2*} \\
        \vdots \\
        \bm{b}_{jm,j-1\,*} \\
        \bm{b}_{jm,j+1\,*} \\
        \vdots \\
        \bm{b}_{jm,N*}
    \end{bmatrix},
\end{align}
where $\bm{a}_{mj,i*}$ and $\bm{b}_{mj,i*}$ are defined as follows,
\begin{align}
    \bm{a}_{jm,i*}&=\begin{cases}
        \bm{q}_{i*}^{(-j)}, & i \notin \mathcal{V}_{jm}, \\
        \bm{q}_{i*}^{(-j)}-\mu \bm{e}_i^{\intercal}, & i \in \mathcal{V}_{jm},
    \end{cases} \label{eq:am} \\
    \bm{b}_{jm,i*}&=\begin{cases}
        q_{ij} \bm{e}_j^{\intercal}, & i \notin \mathcal{V}_{jm}, \\
        q_{ij} \bm{e}_j^{\intercal}+\mu \bm{e}_i^{\intercal}, & i \in \mathcal{V}_{jm}.
    \end{cases} \label{eq:bm}
\end{align}

We now present the following example by fixing $N=5$ with focus on cycle-$1$ and the threshold vector satisfying $0<\tau_{14}<\tau_{12}=\tau_{15}<\tau_{13}<\infty$. Then, we can express $\gamma_{1m}$ and $\mathcal{V}_{1m}$ for given $m$ as,
\begin{alignat}{4}
    \gamma_{11}&=0, \quad \gamma_{12}&&=\tau_{14}, \quad \gamma_{13}&&=\tau_{12}=\tau_{15}, \quad \gamma_{14}&&=\tau_{13}, \quad \gamma_{15}=\infty,\\
    \mathcal{V}_{11}&=\emptyset, \quad \mathcal{V}_{12}&&=\{4\}, \quad \mathcal{V}_{13}&&=\{2,4,5\}, \quad \mathcal{V}_{14}&&=\{2,3,4,5\}. 
\end{alignat}
Now, consider region~$3$, where the threshold values for states $2, 4, 5$ are exceeded, thus transmission is initiated if the process is only in these states. By using \eqref{eq:am} and \eqref{eq:bm}, we can write $\bm{A}_{13}$ and $\bm{B}_{13}$ as,
\begin{align}
    \bm{A}_{13}=\begin{bmatrix}
        -\sigma_2-\mu & q_{23} & q_{24} & q_{25} \\
        q_{32} & -\sigma_3 & q_{34} & q_{35} \\
        q_{42} & q_{43} & -\sigma_4-\mu & q_{45} \\
        q_{52} & q_{53} & q_{54} & -\sigma_5-\mu
    \end{bmatrix} , \quad 
    \bm{B}_{13}=\begin{bmatrix}
        q_{21} & \mu & 0 & 0 & 0 \\
        q_{31} & 0 & 0 & 0 & 0 \\
        q_{41} & 0& 0 &\mu & 0  \\
        q_{51} & 0& 0 & 0 & \mu  
    \end{bmatrix}. \nonumber
\end{align}

Returning to the analysis, we can find $a_j(\bm{\tau}_j)$ and $d_j(\bm{\tau}_j)$ by \eqref{eq:ADij} and using the expression for the first and the second moments given in \eqref{eq:first_moment_ns} and \eqref{eq:second_moment_ns} as,
\begin{align}
    d_j(\bm{\tau}_j)&=\mathbb{E}\left[D_j(\bm{\tau}_j)\right]=\sum_{m=1}^{M-1} I_{1}(\bm{\beta}_{jm},\bm{A}_{jm},\gamma_{jm},\gamma_{j\,m+1})+ I_{1,\infty}(\bm{\beta}_{jM},\bm{A}_{jM},\gamma_{jM}) +\dfrac{1}{\sigma_j},\label{d_ij} \\
    a_j(\bm{\tau}_j)&=\mathbb{E}\left[A_j(\bm{\tau}_j)\right]=\dfrac{1}{2}\sum_{m=1}^{M-1} I_{2}(\bm{\beta}_{jm},\bm{A}_{jm},\gamma_{jm},\gamma_{j \, m+1})+ \dfrac{1}{2}I_{2,\infty}(\bm{\beta}_{jM},\bm{A}_{jM},\gamma_{jM}).  \label{a_ij}
\end{align}

Similarly, we can calculate the transition probability from synchronization point $S_j$ to $S_i$, which is equivalent to the absorption probability to absorption state $S_i$ for the process $Y_j(t)$. By using \eqref{eq:probij}, these probabilities can be written as,
\begin{align}
    p_{ji}(\bm{\tau}_j)&=\sum_{m=1}^{M-1}I_0(\bm{\beta}_{jm},\bm{A}_{jm},\bm{b}_{jm,*i},\gamma_{jm},\gamma_{j\,m+1}) + I_{0,\infty}(\bm{\beta}_{jM},\bm{A}_{jM},\bm{b}_{jM,*i},\gamma_{jM}),
\end{align}
where $\bm{b}_{jm,*i}$ is the $i$th column of $\bm{B}_{jm}$.

The only remaining CSMDP parameter is $c_j(\bm{\tau}_j)$ which corresponds to the expected number of transmissions initiated during cycle-$j$. Let us first define $c_{ji}(\bm{\tau}_j)$ which is the total number of transmissions in  state $i$ for cycle-$j$, and $c_j(\bm{\tau}_j)=\sum_{i\neq j}c_{ji}(\bm{\tau}_j)$. It equals the number of transitions of $X(t)$ from any state to state $i$ in the interval $[\tau_{ji},\infty)$, plus the probability that $X(t)=i$ when the threshold $\tau_{ji}$ is reached. If the region starts with boundary value $\tau_{ji}$ defined as $m_i$, $c_{ji}(\bm{\tau}_j)$ can be written as,
\begin{align} \label{eq:c_ji}
    c_{ji}(\bm{\tau}_j)&=g_{ji}(\tau_{ji})+\sum_{i'\neq i,j} \int_{\tau_{ji}}^{\infty}q_{i'i}g_{ji'}(t)\dd{t}, \\
    &=g_{ji}(\tau_{ji})+\sum_{m=m_i}^{M-1} I_0(\bm{\beta}^{(-i)}_{jm},\bm{A}^{(-i)}_{jm},\bm{q}^{(-j,i)}_{*i},\gamma_{jm},\gamma_{j\,m+1})+I_{0,\infty}(\bm{\beta}^{(-i)}_{jM},\bm{A}^{(-i)}_{jM},\bm{q}^{(-j,i)}_{*i},\gamma_{jM}), 
\end{align}
where $\bm{\beta}^{(-i)}_{jm}$ is $\bm{\beta}_{jm}$ with its $i$th element removed, $\bm{A}^{(-i)}_{jm}$ is the $i$th row and column removed version of $\bm{A}_{jm}$, and $\bm{q}^{(-j,i)}_{*i}$ is the $i$th column of $\bm{Q}$ excluding the $j$th and $i$th terms. 

Finally, we conclude this section with our main result for the optimality of ESAT policy.

\begin{theorem} \label{lem:1}
    The optimum transmission policy for the optimization problem in \eqref{Opt1} for $N\geq2$  is a multi-threshold policy represented by the quantities $\tau_{ji}$ which ensure that the source transmits when $\text{AoII}(t)$ exceeds $\tau_{ji}$ when $\hat{X}(t)=j$ and ${X}(t)=i$ for $i\neq j$, and stays idle when $\hat{X}(t) =  {X}(t)$. Moreover, transmitted packets are preempted when $X(t)$ makes a state transition before the reception of the transmitted packet. 
\end{theorem}

\begin{Proof}
First, it is clear that when $X(t)=\hat{X}(t)$, transmitting a packet has no effect except the additional sampling cost incurred by redundant transmissions. Therefore, we are only interested in the cases when $X(t)\neq\hat{X}$(t). For the case $X(t)=i$, $\hat{X}(t)=j$, $\text{AoII}(t)=\Delta$ and given any $\lambda$, the decision rule $g^\lambda_{ij}(\Delta)$ that indicates whether the transmission is optimum or not is derived in Appendix~\ref{app:lem1}. For any $\Delta'>\Delta$, it is shown in Appendix~\ref{app:lem1} that the difference between the decision rules $g^\lambda_{ij}(\Delta')-g^\lambda_{ij}(\Delta)$ is non-negative, i.e.,
\begin{align}
g^\lambda_{ij}(\Delta')-g^\lambda_{ij}(\Delta) & \geq 0.
\end{align}
Notice that this inequality holds for any $\lambda$ including $\lambda^*$ that finds the optimum policy for the constrained problem. Therefore, if initiating a transmission is optimum when $X(t)=i$, $\hat{X}(t)=j$ and AoII$(t)=\Delta$, it is also optimum for any AoII value greater than $\Delta$. In other words, there exist optimum thresholds $\tau_{ij}$ for starting transmission for $i \neq j$, and the preemption is never optimum unless $X(t)$ changes.
\end{Proof}

\section{Sub-Optimal Transmission Policies} \label{sec:relaxed}
In the previous section, we showed that ESAT policy is optimum for a general CTMC source. However, ESAT requires the computation of $N(N-1)$ parameters, and the complexity of the analysis increases with the number of  regimes $M$. In addition, in the policy iteration step, \eqref{eq:pol} requires an exhaustive search of $N-1$ thresholds, which becomes infeasible for large values of $N$. Therefore, in this section, we consider two sub-optimal policies which are obtained by relaxing the action space of ESAT. The first sub-optimal policy is EAT (estimation aware transmission) in which the optimum threshold is only based on the estimation value. Therefore, the number of thresholds to compute is reduced to $N$, and moreover, at each policy iteration, only a single optimal threshold value is obtained. The second sub-optimal policy is ST (single threshold) policy in which there is only a single system-wide threshold regardless of the estimation and state values, and the optimum value of this threshold is found by line search without a need for a CSMDP formulation. Trading off optimality and complexity, we will show that we can find the related thresholds for CTMCs with relatively larger numbers of states using the relaxed problem formulations. Finally, we note that these relaxed policies provide optimum transmission policies for certain sub-cases.

\subsection{Estimation Aware Transmission (EAT) Policy  } \label{sec:eat}
In this policy, a single threshold value $\tau_j$ is obtained for each \emph{estimated} DTMC state $j$ at the monitor, and the transmission starts when this threshold value is exceeded regardless of the value of $X(t)$. In other words, EAT employs a relaxed action space from ${\rm I\!R}^{N-1}_+$ to ${\rm I\!R}_+$. Hence, the EAT action $\tau_j$ is equivalent to ESAT action $\tau_j \bm{1}$ in cycle-$j$. 

In order to obtain the CSMDP parameters for this policy, we can adapt the calculations of ESAT for $\tau_{ji}=\tau_j$, $\forall j\neq i$. More specifically, for cycle-$j$, the AMC process $Y_j(t)$ has two regimes with a single boundary $\gamma_j=\tau_j$ for the non-zero threshold. Moreover,  the sets corresponding to the two regimes are $\mathcal{V}_1^{(j)}=\emptyset$, and $\mathcal{V}_2^{(j)}=\mathcal{N}\backslash j$. Thus, initial probabilities $\bm{\beta}_{jm}$, the matrix $\bm{A}_{jm}$, and rows of matrix $\bm{B}_{jm}$ and expressed in \eqref{eq:beta1}-\eqref{eq:betam}, \eqref{eq:am}-\eqref{eq:bm} can be written in a more compact way as,
\begin{alignat}{3}
     \bm{A}_{j1}&=\bm{Q}^{(-j)}, \qquad \bm{A}_{j2}&=\bm{Q}^{(-j)}-\mu \bm{I}, \qquad \bm{\beta}_{j1}&=\dfrac{1}{\sigma_j}\bm{q}^{(-j)}_{j*} \label{eq:eat1} \\
     \bm{b}_{j1,i*}&= q_{ij} \bm{e}_j^{\intercal},
    \qquad \bm{b}_{j2,i*}&= q_{ij} \bm{e}_j^{\intercal}-\mu \bm{e}_i^{\intercal}, \qquad \bm{\beta}_{j2}&=\bm{\beta}_{j1} \mathrm{e}^{\bm{A}_{j1} \tau_j}. \label{eq:eat2} 
\end{alignat} 
Additionally, the CSMDP parameters $d_j(\tau_j)$ and $a_j(\tau_j)$ can be obtained from \eqref{d_ij}-\eqref{a_ij} as,
\begin{align} \label{eq:d_j}
    d_j(\tau_j)=& \bm{\beta}_{j1} \mathrm{e}^{\bm{A_{j1}\tau_j}}\bm{M}_{j1}\bm{1}-\bm{\beta}_{j1}\bm{A}_{j1}^{-1}\bm{1}+\dfrac{1}{\sigma_j},  \\ \label{eq:a_j}
    a_j(\tau_j)=& \tau_j \bm{\beta}_{j1} \mathrm{e}^{\bm{A_{j1}\tau_j}}\bm{M}_{j1}\bm{1}-\bm{\beta}_{j1} \mathrm{e}^{\bm{A_{j1}\tau_j}}\bm{M}_{j2}\bm{1}+\bm{\beta}_{j1}\bm{A}_{j1}^{-2}\bm{1}, 
\end{align}
where $\bm{M}_{j1}=\bm{A}_{j1}^{-1}-\bm{A}_{j2}^{-1}$ and $\bm{M}_{j2}=\bm{A}_{j1}^{-2}-\bm{A}_{j2}^{-2}$.

For the EAT policy, the sensor stays idle during regime $1$, and all transmissions are initiated in regime $2$. Therefore, instead of separately calculating  $c_{ij}(\tau_j)$ as in \eqref{eq:c_ji}, $c_j(\tau_j)$ can be calculated directly by finding the total number of state changes in regime $2$ as,
\begin{align}
    c_j(\tau_j)&=\bm{\beta}_1^{(j)} \mathrm{e}^{\bm{A}_{j1}\tau_j} \bm{F}_j \bm{1}, \label{eq:rj}
\end{align}
where $\bm{F}_{j}=(\bm{I}-\bm{D}_{j})^{-1}$ is the characteristic matrix for the second regime, such that the $(m, n)$th term of the matrix $\bm{F}_j$ gives the average number of visits to the $n$th state before absorption, provided the process starts from the $m$th state, and $\bm{D}_{j}$ is the probability transition matrix of the DTMC obtained at the embedded epochs of state transitions of the generator $\bm{A}_{j2}$, which can be obtained according to \eqref{eq:diag}.

The only remaining CSMDP parameter is the transition probabilities among the states. Remember that transition from state $j$ to any state $i\neq j$ is incurred only at the second regime. Thus, we can calculate these probabilities from \eqref{eq:prob} as,
\begin{align} \label{eq:p_j}
    p_{ji}(\tau_j)=&-\bm{\beta}_{j2}\bm{A}_{j2}^{-1}\bm{B}_{j2}\bm{e}_i, \\
    =&-\mu\bm{\beta}_{j1}\mathrm{e}^{\bm{A}_{j1}\tau_j}\bm{A}_{j2}^{-1}\bm{e}_i, 
\end{align}
since $\bm{B}_{j2}\bm{e}_i=\mu\bm{e}_i$ and $\bm{\beta}_{j2}=\bm{\beta}_{j1}\mathrm{e}^{\bm{A}_{j1}\tau_j}$. Additionally, $p_{jj}(\tau_j)=1-\sum_{i\neq j} p_{ji}(\tau_j)$.

Similar to the ESAT policy, the optimum thresholds and the Lagrange multiplier are obtained with the policy iteration algorithm and line search. The main difference of this case is that the expression in Step~3 of Algorithm~\ref{alg:cap_ij} is simplified as,
\begin{align}
      \underset{\tau_j \in {\rm I\!R}_+}{\arg \min}& \quad   f(\tau_j)= \underset{\tau_j \in {\rm I\!R}_+}{\arg \min}\quad\dfrac{\bm{\beta}_{j1} \mathrm{e}^{\bm{A}_{j1}\tau_j} \left( \tau  \bm{M}_{j1}-\bm{M}_{j2}+\lambda \bm{F}_j - \mu\bm{A}_{j2}^{-1}\bm{V}_{j} \right)\bm{1}+k_a }{\bm{\beta}_{j1}\mathrm{e}^{\bm{A}_{j1}\tau_j} \bm{M}_{j1} \bm{1}+k_d} \label{f_j}, 
\end{align}
where $k_a=-\bm{\beta}_{j1}\bm{A}_{j1}^{-2}\bm{1}$, $k_d=-\bm{\beta}_{j1}\bm{A}_{j1}^{-1}\bm{1}+\frac{1}{\sigma_j}$, and $\bm{V}_{j}$ is a $(N-1)\times(N-1)$ diagonal matrix whose diagonal elements are $V_m-V_j, \ m\neq j$. Therefore, the optimum policy can be obtained by line search.

The whole procedure for EAT is involved but it can be simplified in certain cases. In the following, we discuss the specific case of \emph{binary sources} for which closed-form expressions are presented for the CSMDP parameters for each of the two states, and it is shown that EAT is the optimum policy for this specific case.

\begin{definition}
    A binary source is a CTMC with two states and a generator of the form, 
    \begin{align} \label{eq:Q_binary}
        Q=\begin{bmatrix}
        -\sigma_1 & \sigma_1 \\ \sigma_2 & -\sigma_2
    \end{bmatrix}.
    \end{align}
\end{definition}

\begin{remark} \label{rem:N2}
    For the case of binary sources, there are only two thresholds corresponding to $(X(t)=1,\hat{X}(t)=2)$ and $(X(t)=2,\hat{X}(t)=1)$, which makes the EAT policy equivalent to the ESAT policy, and therefore, the optimum policy for binary sources. Closed-form expressions for MAoII and $R$ given the two thresholds, for binary sources, are presented in Appendix~\ref{app:binary}.
\end{remark}

\subsection{Single Threshold (ST) Policy} \label{sec:st}
The final policy proposed in this paper is the ST policy in which the source initiates a transmission when AoII exceeds a single system-wide threshold $\tau$. ST policy can be considered as a special case of ESAT (resp. EAT) with the relaxation of $\bm{\tau}_j=\tau \bm{1},$ $j\in\mathcal{N}$ (resp. $\tau_j=\tau$, $j\in\mathcal{N}$). The main advantage of this policy is that the optimum threshold can be obtained by line search without having to employ a CSMDP formulation. However, the embedded DTMC approach is still used for the analysis of the system for any target threshold value $\tau$. More specifically, MAoII and $R$ can be obtained by substituting $\tau_j=\tau$, $j\in\mathcal{N}$ from the equations for EAT in Subsection~\ref{sec:eat}. The whole procedure for ST is detailed in Section~\ref{sec:imp}. 

Similar to EAT, closed-from expressions can be obtained for the ST policy for the specific case of \emph{symmetric sources} for which ST policy is optimum.

\begin{definition}
     For a symmetric $N$-state CTMC, all state holding times are equal to each other, i.e., $\mathbb{E}[H_j]=\frac{1}{\sigma}$,  $j\in \mathcal{N}$, and all state transition rates are equal.
\end{definition}

From the definition above, the generator matrix $\bm{Q}$ of a symmetric $N$-state CTMC is in the following form,
\begin{align} \label{eq:QSym}
    \bm{Q}=\begin{bmatrix}
        -\sigma & \frac{\sigma}{N-1} & \dots & \frac{\sigma}{N-1} \\
        \frac{\sigma}{N-1} & -\sigma & \dots & \frac{\sigma}{N-1} \\
        \vdots & & \ddots  & \vdots \\
        \frac{\sigma}{N-1} & \dots & \dots & -\sigma
    \end{bmatrix}. 
\end{align}

\begin{remark} \label{rem:HN}
    Because of the symmetry among the states in symmetric CTMCs, ESAT reduces to ST in this case.  Closed-form expressions for MAoII and $R$ for a symmetric source are presented in Appendix~\ref{app:sym}.
\end{remark}

\section{Numerical Results} \label{sec:nr}
In this section, we present our numerical results which are obtained using both analytical models and simulations for the following purposes: i) validate our analysis, ii) showcase the optimality of the proposed transmission policies, and iii) comparatively evaluate the three proposed transmission policies as well as an AoII-agnostic benchmark policy called Poisson sampling (PS) which refers to a transmission policy when the sensor samples the source process with fixed Poisson intensity $\gamma$ when the source and the monitor are out-of-sync, i.e., when $X(t) \neq \hat{X}(t)$. The value of $\gamma$ meeting the budget requirement $b$ is found by line search using the analytical expression of the push-based model proposed in \cite{cosandal2024modelingC} for PS. Simulation results depict the average performances over at least $10^5$ cycles. For all simulations except for Fig.~\ref{fig:sym_th}, optimum policies are calculated analytically, and the simulations are performed for the obtained optimum policies. Specifically, in Fig.~\ref{fig:sym_th}, we provide simulation results for each studied threshold value which is not necessarily optimum. 

\begin{algorithm}[t]
    \caption{Finding the Lagrangian coefficient with bisection search}\label{alg:lag_j}
    \begin{algorithmic}
    \State \textbf{Require:} $b$, $\lambda_{\max}$, $\epsilon_{\lambda}$
    \State \textbf{Initialize:} Initiate $\lambda_\text{min}=0$, and obtain $\phi_0$ by applying Algorithm \ref{alg:cap_ij} for ESAT, or from the adapted version for EAT with $\lambda=0$.
    \If{$R^{\phi_0}\leq b$}
        \State $\phi^*\gets\phi_0$
        \Else
        \State $\lambda_{\text{mid}}\gets \frac{\lambda_{\text{max}}+\lambda_{\text{min}}}{2}$
        \While{$|R^\phi-b|>\epsilon_{\lambda}$}
        \State Apply Algorithm \ref{alg:cap_ij} for ESAT, or from the adapted version for EAT with $\lambda_{\text{mid}}$ and obtain $\phi$.
        \If{$R^\phi\geq b$}
        \State $\lambda_{\text{min}}\gets\lambda_{\text{mid}}$
        \Else
        \State $\lambda_{\text{max}}\gets\lambda_{\text{mid}}$
        \EndIf
        \State $\lambda_{\text{mid}}\gets\frac{\lambda_{\text{max}}+\lambda_{\text{min}}}{2}$
        \EndWhile
    \EndIf
    \State \textbf{Return:} $\phi^*\gets\phi$  
    \end{algorithmic}
\end{algorithm}

\subsection{Implementation} \label{sec:imp}
In this subsection, we provide the details of our algorithm preferences while obtaining our numerical results.

\subsubsection{Optimum Lagrangian Coefficient for ESAT and EAT}
As discussed in Section~\ref{sec:esat}, the optimum Lagrangian coefficient is located at the boundary of the constraint set which is convex. Thus, the optimum coefficient $\lambda^*$ can be obtained by bisection search in Algorithm~\ref{alg:lag_j} which is initiated with $\lambda_{\text{min}}=0$ and $\lambda_{\text{max}}=100$, and the sensitivity parameter of this algorithm is set to $\epsilon_\lambda=10^{-3}$.

\subsubsection{Policy Iteration for ESAT}
For the ESAT policy, the minimization in \eqref{eq:pol} is performed by discretizing each threshold with a step size of $\delta_\tau=0.05$, and exhaustive search is performed for each threshold in the interval $[0, 10]$. 

\subsubsection{Policy Iteration for EAT}
In the policy iteration step of the EAT policy, we employ the gradient descent algorithm to minimize \eqref{f_j}.  Since the elements of $\bm{V}$ can have arbitrary values, we cannot be guaranteed about the convexity of $f(\tau_j)$ in \eqref{f_j}. However, in the numerical results, we have consistently observed that it has at most one minimum. The reason can be understood by the following. Note that the derivative of $f(\tau_j)$ can be written as,
\begin{align}
f'(\tau_j)=\dfrac{\bm{\beta}_{j1}\mathrm{e}^{\bm{A}_{j1}\tau_j}\left( \mathrm{e}^{\bm{A}_{j1}\tau_j} \bm{M}_{j1}^2+k_d \tau_j \bm{A}_{j1}\bm{M}_{j1} +\bm{M}_{j0}   \right)\bm{1}}{\left(\bm{\beta}_{j1}\mathrm{e}^{\bm{A}_{j1}\tau_j} \bm{M}_{j1} \bm{1}+k_d\right)^2},
\end{align}
where $\bm{M}_{j0}$ is the collection of matrices not depending on the value of $\tau_j$, and is written as, 
\begin{align}
\bm{M}_{j0}=k_d \bm{M}_{j1}-k_d\bm{M}_{j2}\bm{A}_{j1}+k_d\lambda\bm{A}_{j1}\bm{F}_{j}-k_d\mu\bm{A}_{j1}\bm{A}_{j2}^{-1}\bm{V}-k_a\bm{A}_{j1}\bm{M}_{j1}.
\end{align}
First, observe that the denominator of $f(\tau_j')$ is the square of the expected duration, hence it is always positive and the expression is definite. Additionally, the term $\bm{\beta}_{j1}\mathrm{e}^{\bm{A}_{j1}}$ corresponds to the initial probability of the second regime, thus its elements are positive and it decays exponentially with $\tau_j$. Moreover, the term $\mathrm{e}^{\bm{A}_{j1}\tau_j} \bm{M}_{j1}^2$ itself also decays exponentially with $\tau_j$, and this double decay helps us to approximate it as,
\begin{align}
\mathrm{e}^{\bm{A}_{j1}\tau_j} \bm{M}_{j1}^2 \approx \bm{M}_{j1}^2+\tau_j\bm{A}_{j1} \bm{M}_{j1}^2.
\end{align}
This approximation makes inside of the parenthesis in the numerator a matrix polynomial function in $\tau_j$ with degree $1$, which indicates that $f'(\tau_j)$ generally has at most $1$ root. Therefore, the continuous-valued threshold can be obtained by the gradient descent algorithm, without needing any discretization of the action space.

\subsubsection{Optimization for ST}
Notice that $d_j(\tau)$ and $a_j(\tau)$ in \eqref{eq:d_j}-\eqref{eq:a_j} are monotonically increasing in $\tau$, and $c_j(\tau)$ in \eqref{eq:rj} is monotonically decreasing in $\tau$, which does not directly lead to the monotonicity of $\text{MAoII}(\tau)=\frac{\sum_{j=1}^{N}\pi_j a_j}{\sum_{j=1}^{N}\pi_j d_j}$ and $R(\tau)=\frac{\sum_{j=1}^{N}\pi_j c_j}{\sum_{j=1}^{N}\pi_j d_j}$. However, we have observed from the numerical examples that they are monotonically increasing, and decreasing functions, respectively. Thus, we propose to use the bisection search algorithm in Algorithm~\ref{alg:tau_st} to obtain the optimum threshold of ST.

\begin{algorithm}[t]
    \caption{Finding the optimum threshold for ST policy}\label{alg:tau_st}
    \begin{algorithmic}
    \State \textbf{Require:} $b$, $\tau_{\max}$, $\epsilon_{\tau}$
    \State \textbf{Initialize:} Initiate $\tau_\text{min}=0$.
    \State Obtain parameters $d_j(\tau_\text{min})$, $a_j(\tau_\text{min})$, $c_j(\tau_\text{min})$ and $p_{ji}(\tau_\text{min})$ by substituting $\tau_j=\tau_\text{min}$ to \eqref{eq:d_j}-\eqref{eq:p_j} for each $j \in \mathcal{N}$ and $i \in \mathcal{N}\backslash j$. By using them, calculate steady-state probabilities $\bm\pi$ by \eqref{eq:oneonetranspose}, and MAoII$(\tau_\text{min})$ and $R(\tau_\text{min})$ values from \eqref{eq:age_app}-\eqref{eq:rate_app}.
    \If{$R(\tau_{\text{min}})\leq b$}
    \State $\tau^*\gets 0$
    \Else
    \State $\tau_{\text{mid}}\gets \frac{\tau_{\text{max}}+\tau_{\text{min}}}{2}$
    \While{$|R(\tau_{\text{mid}})-b|>\epsilon_{\tau}$}
    \State Obtain parameters $d_j(\tau_\text{mid})$, $a_j(\tau_\text{mid})$, $c_j(\tau_\text{mid})$ and $p_{ji}(\tau_\text{mid})$ by substituting $\tau_j=\tau_\text{mid}$ 
    \State to \eqref{eq:d_j}-\eqref{eq:p_j} for each $j \in \mathcal{N}$ and $i \in \mathcal{N}\backslash j$. By using them, calculate steady state
    \State probabilities $\bm\pi$ by ~\eqref{eq:oneonetranspose}, and MAoII$(\tau_\text{mid})$ and $R(\tau_\text{mid})$ values from \eqref{eq:age_app}-\eqref{eq:rate_app}.
    \If{$R(\tau_{\text{mid}})\geq b$}
    \State $\tau_{\text{min}}\gets\tau_{\text{mid}}$
    \Else
    \State $\tau_{\text{max}}\gets\tau_{\text{mid}}$
    \EndIf
    \State $\tau_{\text{mid}}\gets\frac{\tau_{\text{max}}+\tau_{\text{min}}}{2}$
    \EndWhile
    \EndIf
    \State \textbf{Return:} $\tau^*\gets\tau_{\text{mid}}$  
    \end{algorithmic}
\end{algorithm}

\subsection{Symmetric Source}
\begin{figure}[t]
    \centering
    \includegraphics[width=0.55\textwidth]{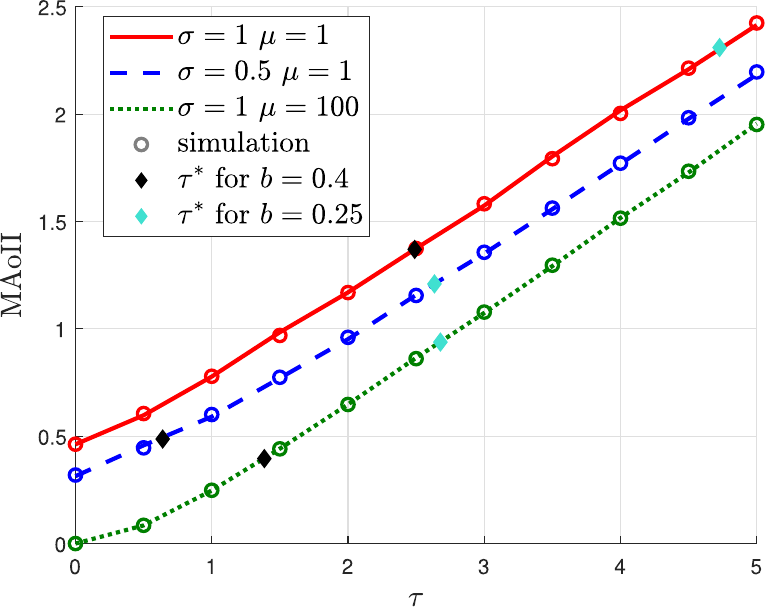}
    \caption{MAoII depicted as a function of the single threshold $\tau$ when $N=20$, and for three values of the pair $(\sigma,\mu)$. Circles are used for simulation results whereas curves are used for analytical results. Optimum threshold values under the sampling constraints $b=0.4$ and $b=0.25$, respectively, are marked by black and turquoise diamonds.}
    \label{fig:sym_th}
\end{figure}

\begin{figure}[th]
    \begin{center}
    \subfigure[$\mu=1$]{\includegraphics[width=0.475\columnwidth]{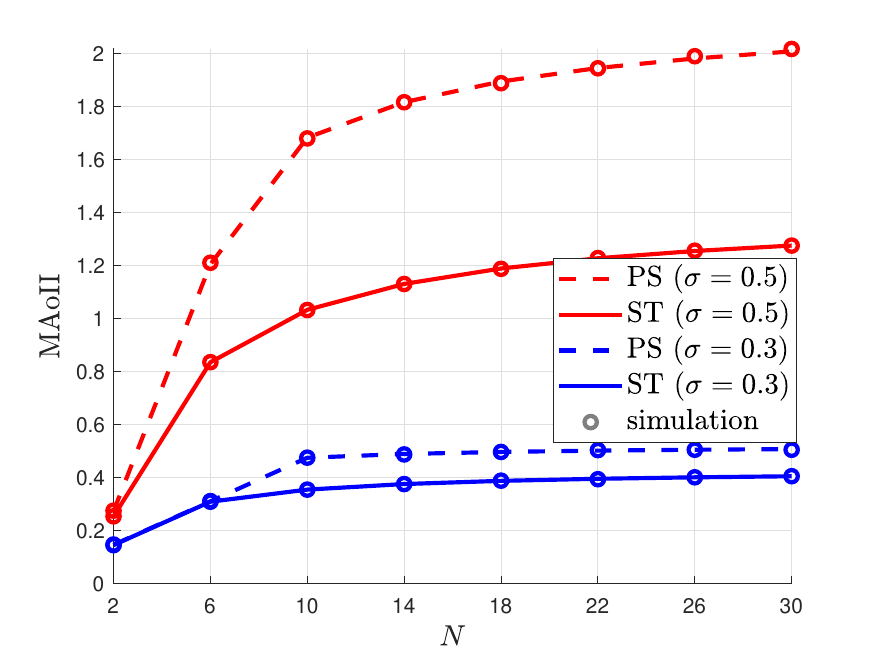}}
    \subfigure[$\mu=100$]{\includegraphics[width=0.475\columnwidth]{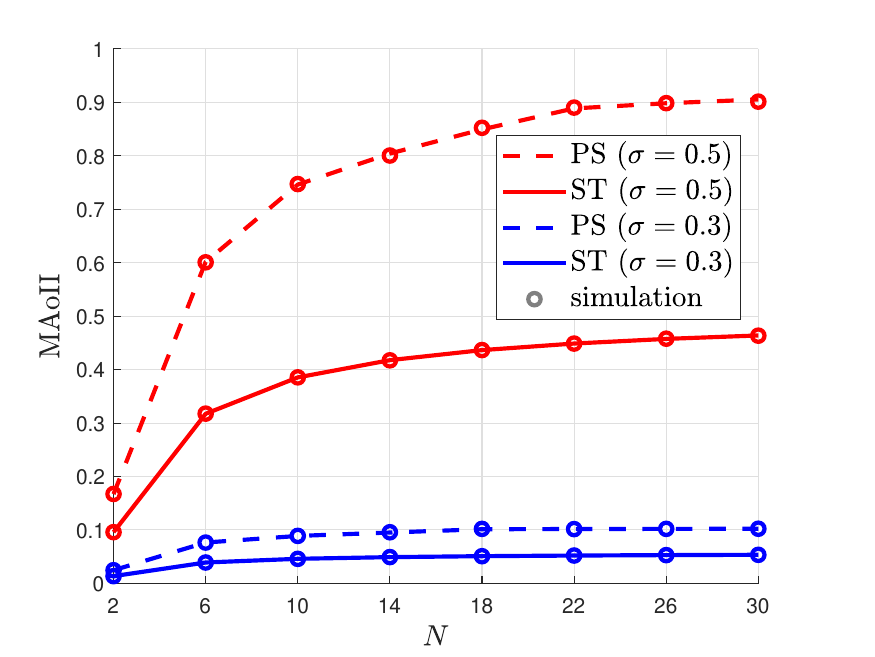}}  
    \end{center}  
    \caption{MAoII depicted as a function of $N$ for ST and PS policies when $b=0.25$ for different service rates. Lines correspond to analytical results, and circles are used for simulations.}
    \label{fig:sym}
\end{figure}

For the symmetric CTMC scenario, we investigate the performance of the optimum ST policy. For this purpose, we construct a CTMC with a generator chosen according to \eqref{eq:QSym}, for given parameters $\sigma$ and $N$. In Fig.~\ref{fig:sym_th}, MAoII is depicted as a function of the single threshold $\tau$, and both analytical and simulation results are given when $N=20$, and for three values of the pair $(\sigma,\mu)$. In Fig.~\ref{fig:sym_th}, circles represent simulation results, and curves represent analytical results. Additionally, the optimum threshold values under the sampling constraints $b=0.4$ and $b=0.25$, respectively, are marked by black and turquoise diamonds. First, observe that for all cases, the function MAoII$(\tau)$ is a monotonically increasing function of $\tau$, and thus, the optimum threshold becomes $\tau_{\min}$ in \eqref{eq:tau_min}. Furthermore, the agreement between analytical and simulation results validates our analysis for symmetric cases. In Fig.~\ref{fig:sym}, the optimum MAoII obtained by the ST algorithm is compared to the PS policy for varying $N$ and for two values of $\sigma$ that make use of the optimum thresholds. Both analytical and simulation results are plotted. We observe that the performance gain of the ST policy compared to PS in terms of MAoII becomes more apparent for larger values of $N$ and for larger $\sigma$. 

\subsection{Binary Source}
\begin{figure}
    \centering
    \includegraphics[width=0.55\textwidth]{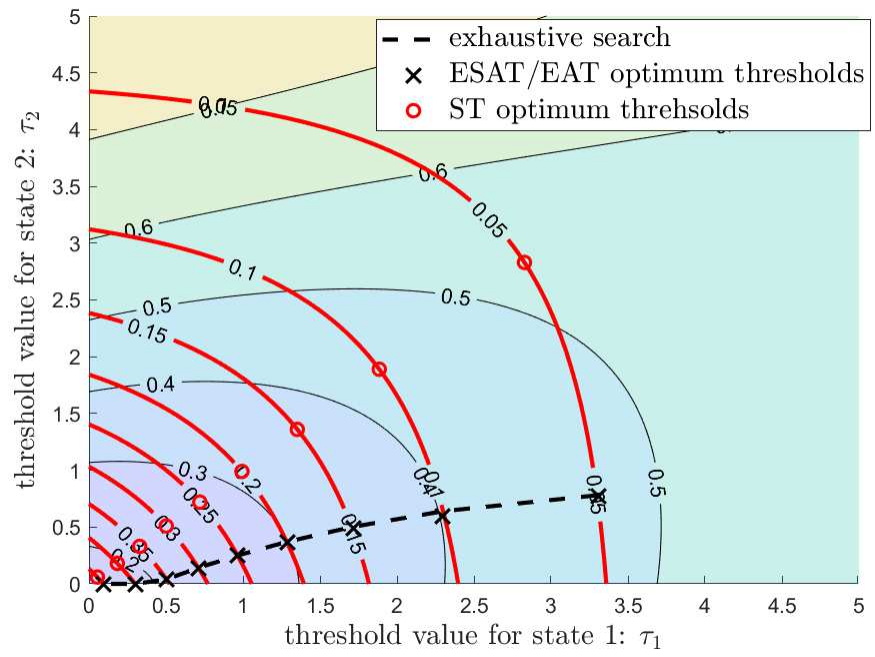}
   \caption{The red (resp. gray) contour lines correspond to the threshold pairs with the same sampling rate $R$ (resp. same MAoII). The minimum MAoII points for a given sampling rate $R$ are connected with a dashed line, whereas the optimum threshold pairs obtained by the proposed EAT and ST policies are marked by a cross and circle, respectively.}
    \label{fig:exh2d}
\end{figure}

For binary sources, ESAT policy reduces to EAT policy as discussed in Remark~\ref{rem:N2} and hence EAT is optimum for this case. In Fig.~\ref{fig:exh2d}, we set $\mu=1$, and plot the contour lines for a binary Markov source with generator given as,
\begin{align}
     \begin{bmatrix}
        -0.6 & 0.6 \\ 0.75 & -0.75
    \end{bmatrix},
\end{align}
for the threshold pair $(\tau_1,\tau_2)$ using the analysis. The red (resp. gray) contour lines correspond to the threshold pairs with the same sampling rate $R$ (resp. same MAoII). The minimum MAoII points for a given sampling rate $R$ are connected with a dashed curve, whereas the optimum threshold pairs obtained by the proposed EAT policy are marked by a cross, which shows perfect agreement between the two, validating the proposed algorithm. Additionally, we mark the optimum threshold for the ST policy that satisfies the budget with red circles. We observe that EAT outperforms ST significantly.
 
\subsection{Ternary Source}
\begin{figure}[t]
    \begin{center}
    \subfigure[$\mu=1$]{\includegraphics[width=0.475\columnwidth]{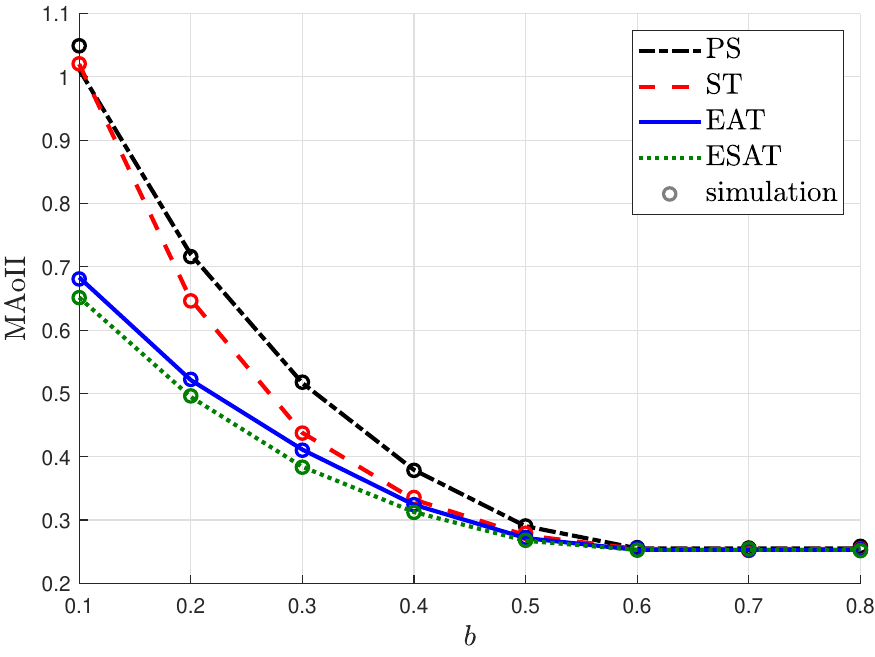}}
    \subfigure[$\mu=100$]{\includegraphics[width=0.475\columnwidth]{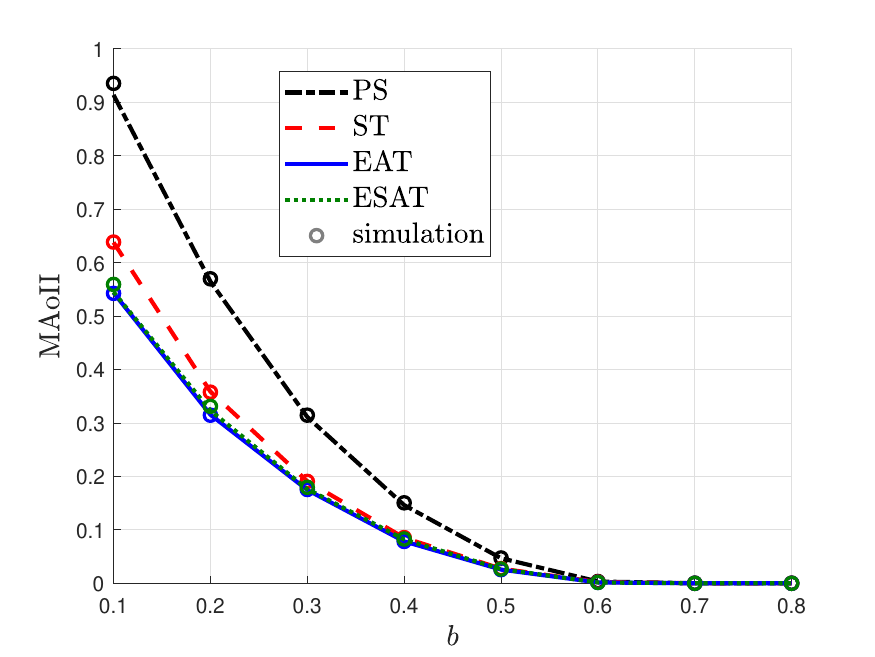}}  
    \end{center}  
    \caption{MAoII depicted as a function of the sampling budget $b$ for the ternary source when the ESAT, EAT, ST, and PS policies are employed for two values of $\mu$. Curves correspond to the analytical results, and simulations are indicated by circles.}
    \label{fig:sym_gen3}
\end{figure}

In this example, we consider a ternary CTMC information source with the generator, 
\begin{align}
  &\begin{bmatrix}
        -1.025 & 1 & 0.025 \\
        0.05 & -0.75 & 0.7 \\
        0.4 & 0.01 & -0.41
    \end{bmatrix}. \label{sources2}
\end{align}
In Fig.~\ref{fig:sym_gen3}, all three proposed policies, namely ESAT, EAT, and ST, are compared along with the PS policy when the sampling budget $b$ is varied, for two values of the service rate $\mu$. Perfect fit with simulations validates the analytical method we use relying on MRPH distributions. As the budget $b$ increases, all policies reduce to the \emph{always transmit} policy. However, the error floor for the $\mu=1$ case stems from moderate packet service times which vanishes when $\mu=100$. Additionally, the performance gaps in the studied policies are more apparent for moderate network delays, i.e., when $\mu=1$, and the ST policy substantially outperforms PS both of which are outperformed by EAT and ESAT. On the other hand, ESAT and EAT perform almost identically when $\mu=100$, i.e., when the channel delay is very low. The underlying reason is that when the channel delay becomes negligible compared to the holding times, i.e., when $\mu<\sigma_j$, $\forall j$, the first initiated transmission in a synchronization state $j$ is almost always successful, thus using multiple thresholds in this particular state becomes less relevant. The slight difference between these two policies stems from the fact that the thresholds for the EAT policy are obtained with a high resolution gradient descent algorithm without discretization which is in contrast to ESAT for which we discretize the action space leading to lesser resolution in finding the optimum thresholds. These simulations are performed by a Windows OS computer with Intel(R) Core(TM) i7-12700K CPU, and 16 GB RAM, and the average computation times taken for the optimization of each policy is summarized in Table~\ref{tab:comp_times}. From the results we obtained for ternary sources, we observe that computation times for PS and ST are of the same order, EAT is two orders of magnitude slower than them, and ESAT is two orders of magnitude slower than EAT.

\begin{table}[t]
    \caption{The average computation times (sec.) for the ternary source.}
    \vspace*{0.2cm}
    \centering
    \begin{tabular}{c|c|c|c|c|} 
     \cline{2-5}
     & PS & ST & EAT & ESAT \\
     \cline{1-5}
     \multicolumn{1}{|l|}{$\mu=1$} &$0.6 \times 10^{-2}$ & $1.21 \times 10^{-2}$ & $1.4$ & $2.2 \times 10^{2} $  \\
     \cline{1-5}
     \multicolumn{1}{|l|}{$\mu=100$} &$ 0.9 \times 10^{-2} $ & $ 2.1 \times 10^{-2} $ & $ 1.4 $ & $ 2.5 \times 10^{2} $  \\
     \cline{1-5}
     \end{tabular}
     \label{tab:comp_times}
\end{table}

\subsection{General $N$-State Source}
\begin{figure}
    \begin{center}
    \subfigure[MAoII vs $N$]{\includegraphics[width=0.475\columnwidth]{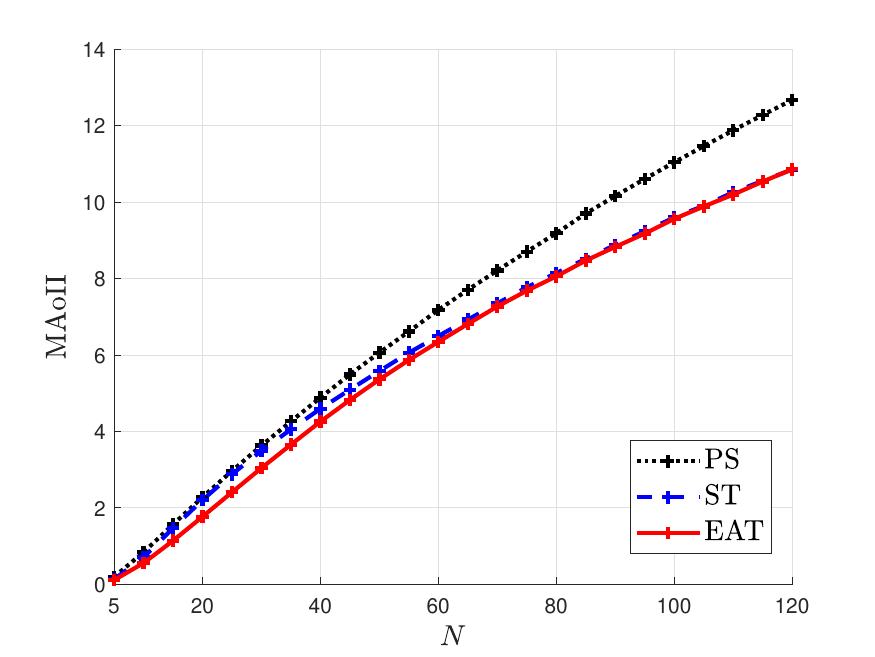}}
    \subfigure[Computational times vs $N$]{\includegraphics[width=0.475\columnwidth]{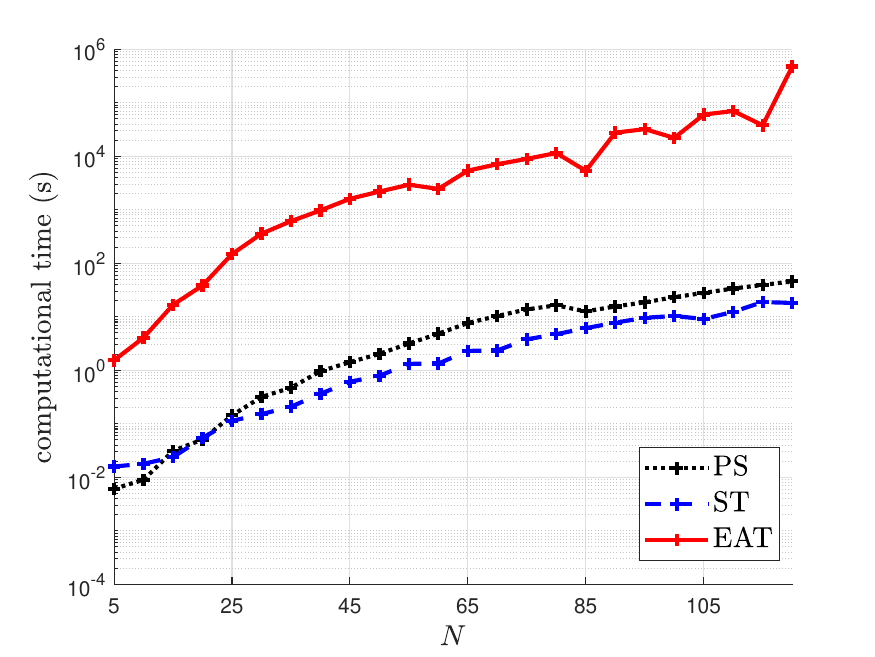}}  
    \end{center}  
    \caption{(a) MAoII, and (b) computation time, depicted as a function of $N$ for the three studied policies PS, ST, and EAT.}
    \label{fig:sym_genLarge}
\end{figure}

In the final example, a comparative evaluation of the EAT, ST, and PS policies is presented in terms of MAoII performance as well as their computation times, for relatively larger state-space CTMC information sources for which ESAT is computationally infeasible. For this purpose, we construct a generator matrix for which the holding rates $q_{ii}=-\sigma_i$ are linearly spread in the closed interval $[\sigma_{\min}, \sigma_{\max}]$ in increasing order, i.e., 
\begin{align}
    \sigma_i = \sigma_{\min}+i \cdot \frac{(\sigma_{\max}-\sigma_{\min})}{N}, \quad i \in \mathcal{N}.
\end{align}
Similarly, transition rates $q_{ij}$, $i \neq j$ are also linearly spread in the interval  $[p_{\min}\frac{\sigma_i}{N-1}, p_{\max}\frac{\sigma_i}{N-1}]$ where ${p_{\min}+p_{\max}}=2$ 
to ensure a legitimate generator. In Fig.~\ref{fig:sym_genLarge}, the problem parameters are set to $\sigma_{\min}=1$, $\sigma_{\max}=20$, $p_{\min}=0.8$, $p_{\max}=1.2$, $\mu=1$, and $b=0.1$ as a representative example for which the behaviors at each state are substantially different.
In Fig.~\ref{fig:sym_genLarge}, we depict the MAoII performance and also the computation times for each of the three policies as a function of the number of states $N$. The observations on computation times are in line with our observations for the ternary source for any value of $N$. In terms of the MAoII performance, when $N$ is varied, ST substantially outperforms PS, and EAT consistently outperforms ST. However, the gap between the latter two diminishes as $N$ increases. 

\section{Conclusions} \label{sec:conc}
In this paper, AoII (age of incorrect information) is investigated for push-based sampling of a CTMC (continuous-time Markov chain), and various AoII-aware threshold-based policies are proposed for the minimization of the average AoII, namely MAoII, under an overall sampling budget. The tool that we use in this paper is the MRPH (multi-regime phase-type) distribution, that we introduce the first time in this paper, to obtain an analytical method to determine the MAoII and also the average sampling rate, for any given age-aware policy. While the multi-threshold ESAT (estimation- and state-aware transmission) policy is shown to be the optimum transmission policy for a general finite-state irreducible CTMC, the complexity of finding the thresholds using exhaustive search increases exponentially with the number of states. Thus, ESAT may not be feasible for CTMCs with large state spaces, for which case we propose two sub-optimum policies with lower complexity, namely, the multi-threshold EAT (estimation-aware transmission) and ST (single threshold) policies, which are shown to be optimum only for special CTMCs and they both rely on well-known line search algorithms. As a benchmark, we compared all the three proposed policies with an AoII-agnostic transmission policy, namely PS (Poisson sampling) which is shown to be substantially outperformed by the AoII-aware ST that has the lowest complexity among the three proposed policies. The multi-threshold policy EAT is about two orders of magnitude slower than ST while we have identified cases, i.e., ternary source, in which the performance gap between EAT and ST is substantial.
Simulation results are also presented to validate the proposed MRPH-based analytical method. Future work consists of more general network delay distributions, two-way delays, packet erasures, metrics other than MAoII, and the use of computationally efficient algorithms, e.g., reinforcement learning (RL), to learn optimum policies even when the dynamics of the CTMC may not be known perfectly in advance. 

\section*{Appendices}

\appendix

\section{Proof of Theorem~\ref{lem:1}} \label{app:lem1}
In order to obtain the structure of the optimum policy, we should re-consider the CSMDP problem in a very general setting for which we consider the source taking actions whenever $X(t)$ or $\hat{X}(t)$ changes, or at infinitesimal time intervals of length $\delta$ if $X(t)$ and $\hat{X}(t)$ are the same during this time interval. The $n$th decision epoch is denoted by $t_n$ at which we denote the state of the system by $s_n=(X(t_n),\hat{X}(t_n),Z(t_n),\text{AoII}(t_n)) \in \mathcal{S}$, with state space
$\mathcal{S}=\{\mathcal{N}\times\mathcal{N}\times\{0,1\} \times {\rm I\!R}_+ \}$. $Z(t_n)$ indicates whether there is an ongoing transmission, i.e., $Z(t_n)=1$, or not, i.e., $Z(t_n)=0$. There are three possible actions $u(t_n) \in \mathcal{U}$ in the action space $\mathcal{U}=\{-1,0,1\}$; $u(t_n)=1$ and $u(t_n)=-1$ correspond to the initiation of a transmission and preemption of an ongoing transmission, respectively, and $u(t_n)=0$ corresponds to ``stay idle" for $Z(t_n)=0$ or ``continue transmission" for $Z(t_n)=1$. Notice that $u(t_n)=1$ (resp. $u(t_n)=-1$) is the only non-zero feasible action when $Z(t_n)=0$ (resp. $Z(t_n)=1$) and if applied, forces the state change $Z(t_n)=1$ (resp. $Z(t_n)=0$).

\begin{figure}[t]
    \begin{center}
    \subfigure[]{\includegraphics[scale=0.95]{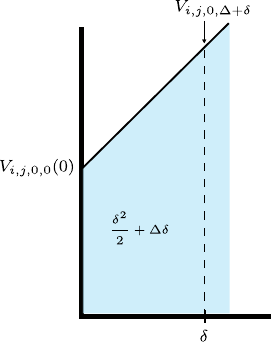}}
    \subfigure[]{\includegraphics[scale=0.95]{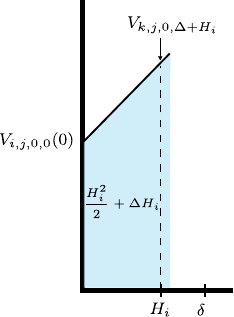}}
    \subfigure[]{\includegraphics[scale=0.95]{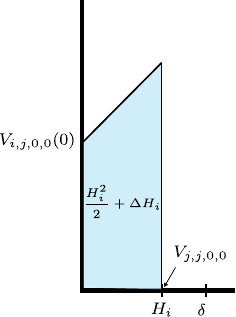}}
    \end{center}  
    \vspace*{-0.4cm}
    \caption{There are three possibilities when action $u=0$ is applied at state $s=(i,j,0,\Delta)$: a) no state change before $\delta$, b) state of observed process changes from $i$ to state $k \neq i,j$ before $\delta$, c) the state of the observed process changes to the estimation before $\delta$. The superscript $\lambda$ is omitted from value functions for convenience.}
    \label{fig:proof0}
\end{figure}

As in Section~\ref{sec:esat}, $D_s(u)$ is the sojourn time when taking action $u$ in state $s$, and $A_s(u)$ and $C_s(u)$  correspond to the area under the $\text{AoII}(t)$ curve and the number of transmission attempts, respectively, during the sojourn time $D_s(u)$. The value of $C_s(u)$ is equal to $1$  when $u=1$, and $0$, otherwise. Again, similar to the formulation given in Section~\ref{sec:esat}, the expected values of these variables are denoted by lowercase letters, and we represent the transition probabilities from state $s$ to $s'$ when action $u$ is applied, with $p_{ss'}(u)$. Finally, for any state $s$ and Lagrange parameter $\lambda$, the Bellman optimality equations in \eqref{eq:opt_gen} can be expressed for this more general formulation as,
\begin{align}
    V_s^\lambda&=\min_{u\in\mathcal{U}} V_s^\lambda(u)=\min_{u\in\mathcal{U}} \left\{a_s(u)+\lambda c_s(u)-\eta d_{s}(u)+\sum_{s' \in \mathcal{S}}^N p_{ss'}(u) V^{\lambda}_{s'}(u)\right\}, \label{eq:opt_gen_app}
\end{align}
where  $V_s^\lambda$ stands for the optimum relative value for state $s$ and $V_s^\lambda(u)$ is the relative value of state $s$ when input $u$ is applied in the current decision interval whereas the optimum policy is applied thereafter. First, it is clear that when $X(t)=\hat{X}(t)$, transmitting a packet has no effect except the additional sampling cost. Therefore, we are only interested in cases when $X(t)\neq\hat{X}$(t). Now, assume we are at state $s=(i,j,0,\Delta)$ for $i \neq j$. If the action $u=0$ is taken, there are three possibilities which are summarized in Table~\ref{tab:u0} and in Fig.~\ref{fig:proof0}, where the Lagrangian cost refers to $A_s(u)+\lambda C_s(u)$. The first case is illustrated in Fig.~\ref{fig:proof0}(a) when the process $X(t)$ does not change, which occurs with probability $\mathbb{P}(H_i>\delta)$ with next state $s'=(i,j,0,\Delta+\delta)$, and $A_s(u)$ equals $\frac{\delta^2}{2}+\Delta\delta$. The second case is illustrated in Fig.~\ref{fig:proof0}(b) when $X(t)$ transitions to state $k \neq j$ before $\delta$, which occurs with probability $\rho_{ik} \mathbb{P}(\delta>H_i)$. In this case, the next state $s'=(k,j,0,\Delta+H_i)$, $\Delta$ increases by $H_i$, and $A_s(u)$ equals $\frac{H_i^2}{2}+\Delta H_i$. The third and final case is illustrated in Fig.~\ref{fig:proof0}(c), where $X(t)$ transitions to state $j$ before $\delta$ with probability $\rho_{ij} \mathbb{P}(\delta>H_i)$. In this case, the value of $\Delta$ returns to $0$, the next state becomes $s'=(j,j,0,0)$, and $A_s(u)$ equals $\frac{H_i^2}{2}+\Delta H_i$. Notice that, for all of these cases, $C_s(u)$ is $0$ because $u=0$. Therefore, we can write $V^{\lambda}_{i,j,0,\Delta}(0)$ as,
\begin{align} 
    V^{\lambda}_{i,j,0,\Delta}(0)= & \mathbb{P}(H_i>\delta) \left( \mathbb{E}\left[\frac{\delta^2}{2}+\Delta \delta\Big|H_i>\delta\right]+ V^{\lambda}_{ i,j,0,\Delta+\delta } -\eta \delta \right) \nonumber  \\
    &+ \rho_{ij}\mathbb{P}(\delta>H_i)\left( \mathbb{E}\left[\dfrac{H_i^2}{2}+\Delta H_i \Big|\delta>H_i\right]+ V^{\lambda}_{ j,j,0 ,0} -\eta H_i \right) \nonumber \\
    &+ \sum_{k\neq j} \rho_{ik} \mathbb{P}(\delta>H_i) \left( \mathbb{E}\left[\dfrac{H_i^2}{2}+\Delta H_i \Big|\delta>H_i\right] + \mathbb{E}\left[V^{\lambda}_{ k,j,0,\Delta+H_i }|\delta>H_i\right]-\eta H_i \right). 
\end{align}
By using the law of total expectation, we can further simplify the above expression to,
\begin{align}    \label{eq:a0short}
    V^{\lambda}_{i,j,0,\Delta}(0)= &\mathbb{P}(H_i>\delta) V^{\lambda}_{ i,j,0,\Delta+\delta}+\rho_{ij} \mathbb{P}(\delta>H_i) V^{\lambda}_{ j,j,0 ,0} \nonumber 
    \\ &+  \sum_{k\neq j} \rho_{ik} \mathbb{P}(\delta>H_i)  
     \mathbb{E}\left[V^{\lambda}_{ k,j,0,\Delta+H_i }\big|\delta>H_i\right]+ \mathbb{E}\left[M_0\left(\Delta-\eta+\frac{M_0}{2}\right)\right], 
\end{align}
where $M_0=\min(H_i,\delta)$.

\begin{table}[t]
\centering
\caption{SMDP parameters for state $s=(i,j,0,\Delta)$, and action $u=0$.} \label{tab:u0}
\vspace*{0.2cm}
\begin{tabular}{|c|c|c|c|} 
\hline
next state & condition & Lagrangian cost & sojourn times\\ \hline
{$(i,j,0,\Delta+\delta)$} & {$H_i>\delta$} & {$\frac{\delta^2}{2}+\Delta \delta$} & {$\delta$} \\ \hline
{$(k,j,0,\Delta+H_i)$}    & {$\delta>H_i, \ i \to k$} & {$\frac{H_i^2}{2}+\Delta H_i$} & {$H_i$} \\ \hline
{$(j,j,0,0)$}             & {$\delta>H_i, \ i \to j$} & {$\frac{H_i^2}{2}+\Delta H_i$} & {$H_i$} \\ \hline
\end{tabular}
\vspace*{0.2cm}
\end{table}

\begin{figure}[t]
    \begin{center}
    \subfigure[]{\includegraphics[scale=0.95]{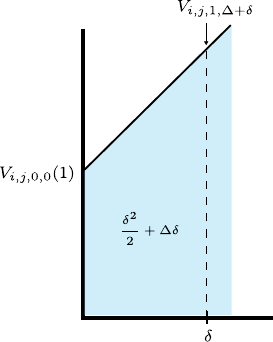}}
    \subfigure[]{\includegraphics[scale=0.95]{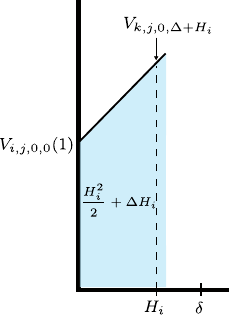}}
    \subfigure[]{\includegraphics[scale=0.95]{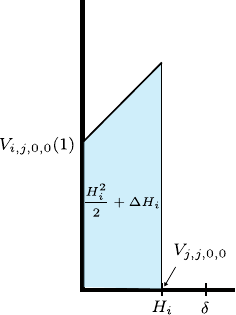}}
        \subfigure[]{\includegraphics[scale=0.95]{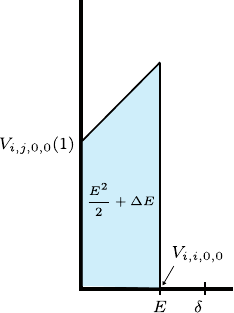}}
    \end{center}  
    \vspace*{-0.4cm}
   \caption{There are four possibilities when action $u=1$ is applied at state $s=(i,j,0,\Delta)$: a) no state change or completed transmission before $\delta$, b) the observed process transitions from $i$ to state $k \neq i,j$ before transmission is complete and $\delta$, c) the state of the observed process changes to the estimated one before transmission is completed and $\delta$, d) transmission is completed before the observed process changes its state and $\delta$. The superscript $\lambda$ is omitted from value functions for convenience.}
   \label{fig:proof1}
\end{figure}

Similarly, we can repeat the same procedure for $u=1$, for which all related parameters are summarized in Table~\ref{tab:u1} and Fig.~\ref{fig:proof1}. Different from $u=0$, if the transmission is not completed, and the state of the process does not change before $\delta$, transmission continues in the next epoch which makes the next state $s'=(i,j,1,\Delta+\delta)$ as illustrated in Fig.~\ref{fig:proof1}(a). Additionally, if the transmission is completed before $\delta$ and the state changes, the value of $\hat{X}(t)$ changes and the next state becomes $s'=(i,i,0,0)$ as illustrated in Fig.~\ref{fig:proof1}(d). Moreover, for all cases, $C_s(u)$ equals $1$. Then, for $s=(i,j,0,\Delta)$ and $u=1$, we have
\begin{align}    \label{eq:a1short}
    V^{\lambda}_{i,j,0,\Delta}(1)=&\ \mathbb{P}(H_i>\delta,E>\delta) \  V^{\lambda}_{ i,j,1,\Delta+\delta} +\rho_{ij} \ \mathbb{P}(E>H_i,\delta>H_i) \ V^{\lambda}_{ j,j,0 ,0} \nonumber  \\
    &+\sum_{k\neq j}\rho_{ik} \ \mathbb{P}(E>H_i,\delta>H_i) \ V^{\lambda}_{ k,j,0,\Delta+H_i } \nonumber \\
    &+\mathbb{P}(\delta>E,H_i>E) \  V^{\lambda}_{ i,i,0 ,0}+\mathbb{E}\left[M_1\left(\Delta-\eta+\frac{M_1}{2}\right)\right] + \lambda,  
\end{align}
where $M_1=\min(H_i,E,\delta)$.

\begin{table}[t]
    \centering
    \caption{ CSMDP parameters for state $s=(i,j,0,\Delta)$, and action $u=1$ } \label{tab:u1}
    \begin{tabular}{|c|c|c|c|} 
    \hline
    next state & condition &  Lagrangian cost &   sojourn times  \\ \hline
    {$(i,j,1,\Delta+\delta)$} & {$H_i>\delta,E>\delta$}  & {$\frac{\delta^2}{2}+\Delta \delta + \lambda$ } & {$\delta$} \\ \hline
    {$(k,j,0,\Delta+H_i)$} &  {$\delta>H_i,E>H_i, \ i \to k$} & {$\frac{H_i^2}{2}+\Delta H_i + \lambda$} & {$H_i$} \\ \hline
    {$(j,j,0,0)$}   & {$\delta>H_i,E>H_i, \ i \to j$} & {$\frac{H_i^2}{2}+\Delta H_i+\lambda$} & {$H_i$}  \\ \hline
    {$(i,i,0,0)$} & {$H_i>E,\delta>E$} & {$\frac{E^2}{2}+\Delta E+\lambda$} & {$E$}  \\ \hline
    \end{tabular}
    \vspace*{0.2cm}
\end{table}

Now, by comparing \eqref{eq:a0short} and \eqref{eq:a1short}, we can define the decision rule $g^\lambda_{ij}(\Delta) = V^{\lambda}_{i,j,0,\Delta}(0) - V^{\lambda}_{i,j,0,\Delta}(1) \gtrless 0$ that indicates whether a transmission is optimum or not according to its sign, for given state $s=(i,j,0,\Delta)$ and $\lambda$ as,
\begin{align}
    g^{\lambda}_{ij}(\Delta)=& \mathbb{E}[M_0(M_0+\Delta-\eta)]-\mathbb{E}[M_1(M_1+\Delta-\eta)] +\mathbb{P}(H_i>\delta) V^{\lambda}_{ i,j,0,\Delta+\delta } \nonumber\\
    &  - \mathbb{P}(H_i,E>\delta) V^{\lambda}_{ i,j,1,\Delta+\delta }
    +\Big(\mathbb{P}(\delta>H_i)-\mathbb{P}(\delta>H_i,E>H_i)\Big)\rho_{ij} V^{\lambda}_{ j,j,0 ,0} \nonumber\\
    & + \sum_{k\neq i,j} \rho_{ik} \mathbb{P}(\delta>H_i) \mathbb{E}[V^{\lambda}_{ k,j,0,\Delta+H_i }|\delta>H_i] - \mathbb{P}(H_i>E,\delta>E) V^{\lambda}_{i,i,0,0} -\lambda \nonumber\\
    & -\sum_{k\neq i,j} \rho_{ik} \mathbb{P}(\delta>H_i,E>H_i) \mathbb{E}[ V^{\lambda}_{ k,j,0,\Delta+H_i }|\delta>H_i,E>H_i]. \label{eq:dec}
\end{align}

We have the following observations:
\begin{itemize}
    \item[1)] $V^{\lambda}_{i,j,0,\Delta+d} \geq  V^{\lambda}_{i,j,1,\Delta+d}$ since the source can preempt an ongoing transmission.
    \item[2)] $\mathbb{P}(\delta>H_i)-\mathbb{P}(\delta>H_i,E>H_i)$ and $\mathbb{P}(H_i>\delta)-\mathbb{P}(E>\delta,H_i>\delta)$ are always positive, and independent of $\Delta$.
    \item[3)] $\mathbb{E}[M_0]-\mathbb{E}[M_1]$ is always positive which is immediate from the definitions of $M_0$ and $M_1$.
    \item[4)] $V^{\lambda}_{i,j,z,\Delta}$ is always non-negative and monotonically non-decreasing with $\Delta$ since the Lagrangian cost is always monotonically non-decreasing with $\Delta$ as can also be seen from Tables~\ref{tab:u0}~and~\ref{tab:u1}.
\end{itemize}

Then, for $\Delta'>\Delta$, we can express $g^\lambda_{ij}(\Delta')-g^\lambda_{ij}(\Delta)$ as,
\begin{align}
    g^{\lambda}_{ij}(\Delta')&-g^{\lambda}_{ij}(\Delta) \nonumber\\
    = & (\Delta'-\Delta)(\mathbb{E}[M_0]-\mathbb{E}[M_1])  +\mathbb{P}(H_i>\delta) (V^{\lambda}_{ i,j,0,\Delta'+\delta }-V^{\lambda}_{ i,j,0,\Delta+\delta }) \nonumber\\
    &-\mathbb{P}(H_i>\delta,E>\delta) \left(V^{\lambda}_{ i,j,1,\Delta'+\delta}-V^{\lambda}_{ i,j,1,\Delta+\delta } \right)\nonumber\\
    &+ \mathbb{P}(\delta>H_i)\sum_{k\neq i,j} \rho_{ik}  \mathbb{E}\left[V^{\lambda}_{ k,j,0,\Delta'+H_i }-V^{\lambda}_{ k,j,0,\Delta+H_i }\big|\delta>H_i\right]   \nonumber\\
    &+ \mathbb{P}(\delta>H_i,E>H_i)\sum_{k\neq i,j} \rho_{ik}  \mathbb{E}\left[V^{\lambda}_{ k,j,0,\Delta'+H_i }-V^{\lambda}_{ k,j,0,\Delta+H_i }\big|\delta>H_i,E>H_i\right]   \\
    \geq & (\Delta'-\Delta)(\mathbb{E}[M_0]-\mathbb{E}[M_1]) \nonumber\\
    &+ \big(\mathbb{P}(H_i>\delta)-\mathbb{P}(H_i>\delta,E>\delta)\big) (V^{\lambda}_{ i,j,0,\Delta'+\delta }-V^{\lambda}_{ i,j,0,\Delta+\delta })\nonumber\\
    &+ \big(\mathbb{P}(\delta>H_i)-\mathbb{P}(\delta>H_i,E>H_i)\big) \sum_{k\neq i,j} \rho_{ik}  \mathbb{E}\left[V^{\lambda}_{ k,j,0,\Delta'+H_i }-V^{\lambda}_{ k,j,0,\Delta+H_i }\big|\delta>H_i\right]  \\
    \geq &  0 
\end{align}
Note that this inequality holds for any $\lambda$ including $\lambda^*$ by which the optimum policy is obtained for the constrained main problem. Therefore, if ``initiating transmission" is optimum when $X(t)=i$, $\hat{X}(t)=j$, and AoII$(t)=\Delta$, it is also optimum for any AoII value larger than $\Delta$. In other words, there exist optimum thresholds $\tau_{ji}$ for $i \neq j$ above which transmission is to be initiated, and preemption is never optimum unless $X(t)$ changes during an ongoing transmission.

\section{Closed-Form Expression for Binary Sources} \label{app:binary}
The analysis of the EAT policy involves square matrices of size $N-1$, rectangular matrices of size $N-1\times N$, and vectors of size $N-1$ in \eqref{eq:eat1}-\eqref{eq:eat2}. For binary sources with $N=2$ and generator given as in \eqref{eq:Q_binary}, these matrix/vector variables of Section~\ref{sec:eat} reduce to either scalars or $ 1 \times 2$ vectors as,
\begin{alignat}{5}
    \bm{A}_{11}=-\sigma_2  , \quad \bm{A}_{12} =-\sigma_2-\mu  , \quad \bm{B}_{11}=\begin{bmatrix} q_{21} & 0  \end{bmatrix}, \quad \bm{B}_{12}=\begin{bmatrix} q_{21} & \mu  \end{bmatrix}, \quad \bm{\beta}_{11}= 1,  \\
    \bm{A}_{21}=-\sigma_1  , \quad \bm{A}_{22}=-\sigma_1-\mu  , \quad \bm{B}_{21}=\begin{bmatrix} q_{12} & 0  \end{bmatrix}, \quad \bm{B}_{22}=\begin{bmatrix} q_{12} & \mu  \end{bmatrix}, \quad \bm{\beta}_{21}= 1, 
\end{alignat} 
from which we present the closed-form expressions for the quantities $d_j(\tau_j)$, $a_j(\tau_j)$, $c_j(\tau_j)$ and $p_{ji}(\tau_j)$ in \eqref{eq:d_j}-\eqref{eq:p_j} as,
\begin{align}
    d_1(\tau_1)& = \dfrac{\mathrm{e}^{-\sigma_2\tau_1}}{\sigma_2+\mu}-\dfrac{\mathrm{e}^{-\sigma_2\tau_1}}{\sigma_2}+\dfrac{1}{\sigma_2}+\dfrac{1}{\sigma_1},  \quad
    d_2(\tau_2)=\dfrac{\mathrm{e}^{-\sigma_1\tau_2}}{\sigma_1+\mu}-\dfrac{\mathrm{e}^{-\sigma_1\tau_2}}{\sigma_1}+\dfrac{1}{\sigma_1}+\dfrac{1}{\sigma_2},  \\
    a_1(\tau_1)& = \tau_1\dfrac{\mathrm{e}^{-\sigma_2\tau_1}}{\sigma_2+\mu}-\tau_1\dfrac{\mathrm{e}^{-\sigma_2\tau_1}}{\sigma_2}-\dfrac{\mathrm{e}^{-\sigma_2\tau_1}}{(\sigma_2+\mu)^2}+\dfrac{\mathrm{e}^{-\sigma_2\tau_1}}{\sigma_2^2}+\dfrac{1}{\sigma_2^2},  \\
    a_2(\tau_2)& = \tau_2\dfrac{\mathrm{e}^{-\sigma_1\tau_2}}{\sigma_1+\mu}-\tau_2\dfrac{\mathrm{e}^{-\sigma_1\tau_2}}{\sigma_1}-\dfrac{\mathrm{e}^{-\sigma_1\tau_2}}{(\sigma_1+\mu)^2}+\dfrac{\mathrm{e}^{-\sigma_1\tau_2}}{\sigma_1^2}+\dfrac{1}{\sigma_1^2},  \\
    p_{12}(\tau_1)& = \dfrac{\mu}{\sigma_2+\mu}\mathrm{e}^{-\sigma_2\tau_1}, \quad p_{11}(\tau_1)=1-\dfrac{\mu}{\sigma_2+\mu}\mathrm{e}^{-\sigma_2\tau_1}, \quad p_{21}(\tau_2) = \dfrac{\mu}{\sigma_1+\mu}\mathrm{e}^{-\sigma_1\tau_2}, \\
    \quad p_{22}(\tau_2) & =1-\dfrac{\mu}{\sigma_1+\mu}\mathrm{e}^{-\sigma_1\tau_2},   \quad
    c_1(\tau_1) = \mathrm{e}^{-\sigma_2\tau_1}, \quad c_2(\tau_2)=\mathrm{e}^{-\sigma_1\tau_2}. 
\end{align}
In addition, for $N=2$, the steady-state probabilities for synchronization values in \eqref{eq:oneonetranspose} are,
\begin{align}
    \bm{\pi}=\begin{bmatrix} \dfrac{p_{21}(\tau_2)}{p_{21}(\tau_2)+p_{12}(\tau_1)} & \dfrac{p_{12}(\tau_1)}{p_{21}(\tau_2)+p_{12}(\tau_1)} \end{bmatrix}.  
\end{align}
Finally, the expressions for MAoII$(\bm\tau)$ and $R(\bm\tau)$ for $\bm\tau=\left[ \tau_1 \ \tau_2\right]$ 
become,
\begin{align}
    \text{MAoII}(\bm{\tau})& = \dfrac{\pi_1 a_1(\tau_1)+ \pi_2 a_2(\tau_2)}{\pi_1 d_1(\tau_1)+ \pi_2 d_2(\tau_2)},   \quad
    R(\bm{\tau})=\dfrac{\pi_1 c_1(\tau_1)+ \pi_2 c_2(\tau_2)}{\pi_1 d_1(\tau_1)+ \pi_2 d_2(\tau_2)}.  
\end{align}

\section{Closed-Form Expression for Symmetric Sources} \label{app:sym}
First, note that, for a symmetric CTMC and for any threshold value, the steady-state probabilities for synchronization values are the same, i.e., $\pi_j=\frac{1}{N}$, $j\in\mathcal{N}$. Then, the expressions of MAoII and $R$ are reduced to MAoII and $R$ expression for a single cycle as,
\begin{align}
    \text{MAoII}(\tau)=\dfrac{a(\tau)}{d(\tau)},  \quad
    R(\tau)=\dfrac{c(\tau)}{d(\tau)},  \label{MAoII_ST}
\end{align}
where $d(\tau)$, $a(\tau)$ and $c(\tau)$ are the expected values of duration, total AoII, and the number of transmissions in a single cycle for threshold $\tau$, respectively, and they are obtained from ~\eqref{eq:d_j}-\eqref{eq:rj} by dropping the index subscripts as,
\begin{align}
    d(\tau)& = \bm{\beta}_1 (\bm{A}_1^{-1}-\bm{A}_2^{-1})\mathrm{e}^{\tau \bm{A}_1}\bm{1}- \bm{\beta}_1 \bm{A}^{-1} \bm{1}+\frac{1}{\sigma}, \\
    a(\tau)& = \tau \bm{\beta}_1\mathrm{e}^{\tau \bm{A}_1}(\bm{A}_1^{-1}-\bm{A}_2^{-1})\bm{1}+\bm{\beta}_1\mathrm{e}^{\tau \bm{A}_1}(\bm{A}_2^{-2}-\bm{A}_1^{-2})\bm{1}+\bm{\beta}_1\bm{A}_1^{-2}\bm{1}, \\
    c(\tau)& = \bm{\beta}_1 \mathrm{e}^{\tau \bm{A}_1} \bm{F} \bm{1}. 
\end{align}
Note that the initial PMF for the first regime is equiprobable for any state. Therefore, 
\begin{align}
    \bm{\beta}_1=\dfrac{1}{N-1}\bm{1}^{\intercal}. 
\end{align}
Furthermore, we can express the remaining matrices compactly for symmetric CTMCs as,
\begin{align}
    \bm{A}_1& = \dfrac{N\sigma}{N-1}\left(\bm{I}-\dfrac{1}{N}\bm{J}_{N-1}\right),  \\
    \bm{A}_2& = \bm{A}_1-\mu \bm{I}=\dfrac{N\sigma-(N-1)\mu}{N-1}\left(\bm{I}-\dfrac{\sigma}{N\sigma-(N-1)\mu}\bm{J}_{N-1}\right),  \\
    \bm{F}& = (\bm{I}-\bm{D})^{-1}=\left(\bm{I}+\dfrac{\sigma}{(N-1)(\sigma+\mu)} \bm{J}_{N-1}\right)^{-1}. 
\end{align}
Now, consider the following matrix identities derived from Sherman–Morrison formula \cite{horn2012matrix} and the properties of the matrix exponential given in \cite{wilcox1967exponential},
\begin{align}
        \left[a \left( \bm{I}+c \bm{J}_K \right)\right]^{-1} & =  a^{-1} \left(\bm{I}-\dfrac{c}{cK+1}\bm{J}_K\right), \label{eq:mat_inv} \\
       \mathrm{e}^{a \left( \bm{I}+c \bm{J}_K \right)}& = \mathrm{e}^a \mathrm{e}^{\bm{I}} \mathrm{e}^{c\bm{J}_K}  =\mathrm{e}^{a+1}\left( \bm{I}+\dfrac{\mathrm{e}^K-1}{K}\bm{J}_K\right). \label{eq:mat_expm} 
\end{align}
By using these two identities, we can write the quantities $c(\tau)$, $d(\tau)$ and $a(\tau)$,  in terms of the problem parameters $N$, $\sigma$, $\mu$ and $\tau$ as,
\begin{align}
    c(\tau)& = \dfrac{\mathrm{e}^{-\frac{{\sigma\tau}}{N-1}}}{1-\frac{\sigma(N-2)}{(\sigma+\mu)(N-1)}},  \\
    d(\tau)& = (1-\mathrm{e}^{-\frac{{\sigma\tau}}{N-1}})\dfrac{N-1}{\sigma}+\mathrm{e}^{-\frac{{\sigma\tau}}{N-1}}\dfrac{N-1}{k_1(k_2N+N-1)}+\dfrac{1}{\sigma}, \\
    a(\tau)& = \tau \mathrm{e}^{-\frac{{\sigma\tau}}{N-1}}\dfrac{N-1}{\sigma}
    -\tau \mathrm{e}^{-\frac{{\sigma\tau}}{N-1}}\dfrac{N-1}{k_1(k_2N+N-1)}\nonumber \\
    & \; \; \;+ \mathrm{e}^{-\frac{{\sigma\tau}}{N-1}}\dfrac{(N-1)^2}{k_1^2(k_2N+N-1)^2}
    -\mathrm{e}^{-\frac{{\sigma\tau}}{N-1}}\dfrac{(N-1)^2}{\sigma^2}+\dfrac{(N-1)^2}{\sigma^2}, 
\end{align}
where $k_1=\sigma N + \mu(N-1)$, and $k_2=\dfrac{\sigma}{\sigma N-\mu(N-1)}$. Using these, MAoII$(\tau)$ and $R(\tau)$ can be obtained from \eqref{MAoII_ST}.

Finally, we express the optimization problem that minimizes the MAoII under a sampling rate constraint for the ST policy as
\begin{mini}
	{{\tau}}{\text{MAoII}({\tau}) }
	{\label{Opt1_ST}}
    {}
	\addConstraint{ R({\tau}) }{\leq b} 
\end{mini}
Note that, R$(\tau)$ is a monotonically decreasing function with $\tau$. Therefore, the optimum threshold should be $\tau^*\geq\tau_{\min}$ such that $\tau_{\min}$ can be obtained by solving the following non-linear equation,
\begin{align}
        R(\tau_{\min})-b=0. \label{eq:tau_min}
\end{align} 

\bibliographystyle{unsrt}
\bibliography{bibl}

\end{document}